\documentclass[preprint,12pt]{elsarticle}
\usepackage{amssymb}
\usepackage{times}
\usepackage{soul}
\usepackage{todonotes}
\usepackage[utf8]{inputenc}
\usepackage{amsmath}
\usepackage{amsfonts}
\usepackage{subfigure}
\usepackage{graphicx}
\usepackage{hyperref}
\usepackage{tabularx}
\usepackage{url}
\usepackage{algorithm}
\usepackage{algorithmic}
\usepackage{multirow}
\usepackage{bm}
\usepackage{amsthm}
\usepackage{physics}
\usepackage{color}

\usepackage{amssymb}
\usepackage[toc,title,page]{appendix}
\usepackage{bbm}

\begin{document}
\begin{frontmatter}
\title{Optimal Market Making in the Chinese Stock Market:  A Stochastic Control and Scenario Analysis}

\author[amss,ucas]{Shiqi Gong}
\address[amss]{Academy of Mathematics and Systems Science, Chinese Academy of Sciences, Zhongguancun East Road, Beijing 100190, China}
\address[ucas]{University of Chinese Academy of Sciences, No.19 Yuquan Road, Beijing 100049, China}
\ead{gongshiqi15@mails.ucas.ac.cn}

\author[tud]{Shuaiqiang Liu}
\address[tud]{Delft Institute of Applied Mathematics, Delft University of Technology,  Mekelweg 4, 2628 CD Delft, The Netherlands}
\ead{s.liu-4@tudelft.nl}

\author[pcl]{Danny D. Sun\corref{cor1}}
\address[pcl]{Pengcheng Laboratory, No.2 Xingke 1st Street, Nanshan District, Shen Zhen 518055, Guangdong, China}
\cortext[cor1]{Corresponding author}
\ead{sundn@pcl.ac.cn}

\begin{abstract}
Market making plays a crucial role in providing liquidity and maintaining stability in financial markets, making it an essential component of well-functioning capital markets. Despite its importance, there is limited research on market making in the Chinese stock market, which is one of the largest and most rapidly growing markets globally. To address this gap, we employ an optimal market making framework with an exponential CARA-type (Constant Absolute Risk Aversion) utility function that accounts for various market conditions, such as price drift, volatility, and stamp duty, and is capable of describing 3 major risks (i.e., inventory, execution and adverse selection risks) in market making practice, and provide an in-depth quantitative and scenario analysis of market making in the Chinese stock market. Our numerical experiments explore the impact of volatility on the market maker's inventory. Furthermore, we find that the stamp duty rate is a critical factor in market making, with a negative impact on both the profit of the market maker and the liquidity of the market. Additionally, our analysis emphasizes the significance of accurately estimating stock drift for managing inventory and adverse selection risks effectively and enhancing profit for the market maker. These findings offer valuable insights for both market makers and policymakers in the Chinese stock market and provide directions for further research in designing effective market making strategies and policies.

\end{abstract}


\end{frontmatter}

\section{Introduction}

In a well-developed stock trading market, the market participants are often categorized into three different types according to the information base and trading purpose \citep{Noise1986,HANIF2014429}.  The first type is the information traders, who make the trading decision based on the fundamental analysis or private information about the expected future price movement and company growth prospects or to optimally re-balance his or her portfolio to achieve desired risk-return. His or her target price may be different from the current market price and thus may often trade in a directional and aggressive manner.  The second type is the noise traders, who lack the information and thus the knowledge of the informed value of the stocks, exhibit more speculative nature and use the noises as the basis for trade. Pure index followers are also noise traders who adjust their positions passively at the index rebalance time. The noise traders as a whole are liquidity traders in the sense that they trade  frequently, causing price oscillations and allowing price to be observed. And the third type, the market makers,  provides liquidity by submitting simultaneously to the stock exchanges the limit-price buy (bid) and sell (ask) orders, which are queued, according to certain price-time priority rules set by the exchanges, in the limit-order book (LOB) and wait passively for the active counter-party market orders or marketable limit orders to fulfill the trade. 

The market makers make a profit from the bid-ask spread while managing the risks by adapting their quotes dynamically.  The first common risk a market maker faces and manages is the inventory risk.  The  inventory risk is determined by the amount of  stocks a trader holds or shorts, which are exposed to the price uncertainty due to the market volatility.  Averse to the market price volatility, a market maker with a net long inventory will adjust his or her quoting spread by bidding conservatively and asking aggressively in order to reduce his or her probability to buy and increase his or her probability to sell, and vice versa.  The second common risk is the execution risk.  A market maker will adjust his or her quotes adeptly in order to gain the priority for his or her limit orders to be fulfilled with counter-parties.   And the third common risk is the adverse selection risk. The counter-parties with the directional price drift knowledge from news or other information pick up  the passive limit orders of the market maker in the direction against the maker. For example, in the case of the mid-quote falling, the limit bid order of the market maker are kept fulfilled by the counter-party, and the re-submissions of the limit bid and ask orders  from the maker simply following the price trend passively will keep losing money. In other words, to avoid the adverse selection risk, a market maker may need to identify and forecast the directional drift of the price and set his or her bid and ask prices proactively.   
              
According to \textsc{WIND}$^{\circledR}$ data, as of December, 2022, with a total value of 73 trillion RMB (10.5 trillion USD dollars) , the Chinese stock market is currently one of the largest stock markets in the world, second only to the US market in size.  However, the Chinese market bears distinctive market features and trading regulations compared to the well-developed markets such as the US one. Currently there are  three centralized stock exchanges in China, i.e., Shanghai, Shenzhen  and Beijing Exchanges. Each stock is exclusively listed and traded in one of the three exchanges, therefore no such rules as National Best Bid Offer (NBBO) exist in China as seen in US.  There are no dark pools as trading venues and the short selling is difficult to carry out \cite{ChinaMarket2022}.\footnote{Bessler and Vendrasco, by comparing stock markets with and without short-selling in European market, found that the ban of short-selling could lower the trading activities (including liquidity and volumes) and make the bid-ask spread wider \cite{Banshortselling2022}.}  Retail investors dominate the Chinese stock market in terms of the trading volume, unlike  developed markets where institutional investors dominates \cite{LI2010448}.  The automated trading, despite rapid growth, is still of a low degree \cite{ChinaMarket2022}.  For example, Liu \cite{HFTChinaLIU2021} found that the stocks with the higher participation of high-frequency trading (HFT) in China manifested higher profitability and the regulation tightening negatively impacted the HFT activities.\footnote{The profit advantage of HFT declined gradually after 2019, which may be due to the higher market competition among growing high frequency traders.}  
The market regulation, advanced technology and market distinct features shape the trading activities in China, attracting strong research interests in academia and practice alike. 

In recent years, the market making practices have been introduced into the secondary capital market in China, and in 2022 the newly formed Beijing Exchange adopted the stock market making with the designated securities firms as the market makers. An effective market making operation usually requires a high-frequency algorithm to submit orders timely and optimally so that the market makers have the opportunities from the frequent transactions to profit from the bid-ask spread and/or obtain commission rebates of the exchange, and at the same time manage the risks. In Chinese stock markets, certain rules and regulations such as the T+1 turnover and the stamp tax on stock selling may hinder the market makers from implementing his or her market making strategies. Thus, facing the growth of the Chinese market making practices, in this paper we are keen to investigate theoretically how the market conditions such as price drift and volatility in normal and stress conditions, and the market rules and regulations such as commission rate and stamp tax, will affect the market maker strategy, e.g., the profit taking and the simultaneous management of the inventory, execution and adverse selection risks, so as to realize the effective market making and liquidity provision.  

Various market making models have been proposed and numerical techniques explored in literature.
Avellaneda and Stoikov \cite{avellaneda2008high} set the optimal market making on the stochastic control base, and with the market price simplified as an exogenous stochastic process, a multi-dimensional Hamilton-Jacobi-Bellman (HJB) equation was established for solving the limit order quotes from the market maker as the controls of his or her cash amount and inventory. 
Using the similar setup as \cite{avellaneda2008high} and using a change of variable technique,  Guéant et. al. \cite{inventory2013} simplified the high-dimensional HJB equation into a linear ordinary differential equation system and  solved the market making strategies under inventory constraints.
Its closed-form approximation was applied into practice in certain conditions according to \cite{Optmarketmaking2017}.  
Cartea et. al. \cite{buylowsellhigh2014} proposed a model accounting for arrival of market orders 
and its effects on the subsequent arrivals of the external market order flows, the limit order fill rate (thus the LOB shape) and the short-term drift of the mid-quote. The authors concluded that to avoid being adversely selected by informed traders a maker should develop short-term-alpha predictors in his or her market making strategies. 
More advanced models, such as stochastic volatility \cite{HestonVol2020}, interaction between the limit and market orders \cite{buylowsellhigh2014}, were also developed.   
Since closed-form solutions to the HJB equation are hardly available in most cases, discretization methods are required to approximate numerically the viscosity solutions  (classical solution may not exist). Labahn and Forsyth \cite{FORSYTH2007} developed a finite difference scheme to numerically solve HJB equations. As a follow-up, Forsyth  \cite{FORSYTH2011241} employed the above scheme to obtain a numerical solution to an HJB equation for the optimal trade execution. Other numerical techniques to solve HJB, especially for high dimensional problems, including Monte Carlo simulations \cite{CONG201623} and  deep neural networks \cite{Han8505}, were also explored.  

The contributions to the literature in our current investigation are in the following four aspects.  First, we adapt the theoretical framework for optimal market making based on the stochastic optimal control proposed by \cite{guilbaud2013optimal}. \footnote{The framework in its general form assumes an exogenous Levy process for mid-quote and an exogenous Markov chain for bid-ask spread, and the bid-ask spread takes multiples of the ticks, capable of describing the discrete-valued nature in real market practice. This framework allows one to seek the optimal market making strategies using the limit and market orders to maximize the terminal cash amount profiting from the bid-ask spread while penalizing the inventory during the course. Numerical experiments with price drift were not carried out in \cite{guilbaud2013optimal}.}  In our current work, we explicitly study the scenario of the non-zero mid-quote price drift, which is the short-term price movement predicted by the maker, to accommodate the adverse selection risk, in addition to the inventory risk and execution risk. Being aware of the limitation of the linear utility function in terms of price drift accommodation\footnote {Two types of utility functions were proposed in \cite{guilbaud2013optimal}.  The first one was the linear utility function, which only accepted the martingale mid-quote price process, and the numerical solutions were presented.  The second one was the CARA utility function, which allowed for a more general drifted diffusion process for the mid-quote.}, we take the CARA form of the utility function accordingly, and use a numerical technique to seek the optimal market making strategy solutions in a variety of market and regulation conditions.  Second, the market maker is allowed to bid and ask inside and outside the best bid and ask levels by one tick to accommodate the price drift in his or her optimal strategies. Third, based on the parameters calibrated from the Chinese stock market, we solve the optimal quotes a market maker should submit. Fourth, we investigate under the regulation in the Chinese market the effect of the market price parameters such as price drift and volatility, and the effect of regulation rule such as the stamp tax rate, on the market makers profits and the market liquidity provision.  This helps simulate the behaviors of the market maker, who would act or not act as an effective market liquidity provider based on his or her risk-return balance.

Our numerical experiments cover three key aspects of market making, including the impact of volatility, the role of stamp duty rate, and the effect of stock drift. The results reveal that volatility exerts an adverse effect on the absolute inventory of the optimal market making strategy, aiming to mitigate inventory risk. Moreover, the stamp duty rate negatively impacts both the profit of the market maker and the liquidity of the market. Also, the total tax paid by the market maker tends to follow a concave function concerning the stamp duty rate. A high stamp duty may decrease the collected stamp tax due to its hinderance to the trading volume from the maker. Additionally, an accurate estimation of stock drift is crucial for the market maker to manage inventory and adverse selection risks. These findings can facilitate market makers in devising their tactics and provide policymakers with insights into modifying market regulations, such as the stamp duty rate.

The remainder of this paper is organized as follows. In Section \ref{sec:method}, the mathematical model of a market making problem considering the limit and market orders is briefly introduced. More details are listed in Appendix.  In Section \ref{sec:num_tec}, the numerical method to solve the stochastic optimal control problem is described. In Section \ref{sec:result}, numerical results are presented. In Section \ref{sec:disscusion} we discuss market makers in China and conclude.

\section{Market Data and Model Framework} \label{sec:method}

\subsection{Market Rules and Stock Data}
Currently, there exist three centralized stock exchanges in China, namely, Shanghai, Shenzhen, and Beijing Exchanges. In this paper, we mainly focus on the Shenzhen Stock Exchange due to its high liquidity.  
The Shenzhen Stock Exchange (SZSE) operates an electronic order book system under a set of key trading rules that foster a transparent and well-regulated market environment. Trading hours are divided into two sessions: the morning session from 9:30 AM to 11:30 AM and the afternoon session from 1:00 PM to 3:00 PM (China Standard Time). To ensure smooth trading, the minimum tick size $\delta$ is set at fixed $0.01$ Yuan, representing the smallest price increment for securities. Investors have the flexibility to place various order types, such as limit and market orders, while short-selling is allowed but subject to specific regulations, including the obligation to borrow shares prior to executing a short sale. To maintain market stability, daily price change limits are enforced, typically $\pm 10\%$ for most stocks and $\pm 20\%$ for ChiNext Board stocks. Lastly, the SZSE follows a T+1 settlement cycle, in which securities can be sold one business day after the trade date, while funds from selling are instantly available for purchasing other stocks.
Within the SZSE, the transaction fee comprises two components: stamp duty and commission fee. The stamp duty, represented by a percentage $\rho$, is collected by the government and solely applied when selling stocks. The commission fee, denoted by a percentage $\varepsilon$, is levied by the brokerage and stock exchange during both buying and selling of stocks. In the current SZSE, the stamp duty rate $\rho$ is fixed at $1$‰, while the commission fee $\varepsilon$ varies by brokerage, with a maximum limit not exceeding $1$‰. Typically, for a market maker, the commission fee ranges between $0.1$‰ and $0.3$‰.

Regarding the data provided by the SZSE, three types of data exist for each stock: snapshot data, order data, and tick data. Snapshot data encompasses the top ten bid/ask price levels along with corresponding volumes at a specified snapshot moment, occurring at a frequency of three seconds. Order data comprises the submission time, type, direction (buy/sell), target volume, and price for each submitted order. Tick data incorporates the trade price, volume, direction (buyer/seller-initiated), type (trade/cancel) of every trade, as well as the two orders culminating in the trade. As high-frequency stock data is required for market making, we reconstruct the order book using order data and tick data and mainly use snapshot data generated at a frequency of 10ms for our study.

\subsection{ Optimal Market Making Strategy }

A limit order is an order to buy (respectively sell) at a specific price or below (respectively above).  When a counter-party market order arrives,  the exchange will match the market order with the best available price of the limit orders in LOB and the sub-optimal prices in turn if the market order is not fully fulfilled. In our study, a market maker trader is assumed to submit limit orders to provide the liquidity, or post market orders to consume market liquidity for immediate execution. We investigate the effect of the trade-off between execution priority and quote price under different market conditions for the market maker. Therefore, in our current setup, a market maker has three price options to select when placing a limit order:  posting to the current best bid (ask) price, posting to one-tick higher bid price (lower ask) price, or posting to one-tick lower bid price (higher ask price), see Figure \ref{fig:LOB}.  Limit orders with higher bid prices (lower ask prices) have higher priority in order execution for the maker, but come with a worse quote compared to the current best bid (ask price). On the other hand, limit orders with lower bid (higher ask) prices have less risk of adverse selection for the maker, but have lower priority in order execution.  In our current setup, the fulfilling of the maker limit orders is described by the filling intensities of his or her limit orders, queueing time of his or her limit orders in LOB is not explicitly modeled.  In parallel, in our study, the market orders from the market makers are assumed sufficiently small in volume and immediately fulfilled at the best counter-party price in LOB. No price impact on market movement from the market order execution is explicitly considered herein.

The optimal market making is formulated as a stochastic optimal control problem in our study. We define  a filtered probability space $(\Omega, \mathcal{F},\mathbb{F}, \mathbb{P})$ for stochastic processes, where  the filtration $\mathbb{F} = (\mathcal{F}_t), {t \geq 0}$ satisfies the usual conditions.  Here  trading occurs within a finite time horizon $0<T<\infty$ (e.g., a trading day).

In this section, we briefly explain our mathematical model setup adapted from \cite{guilbaud2013optimal}. More details of the model setup can be found in Appendix~\ref{app:a}. 

\subsubsection{State Variables} \label{sec:state}

\begin{table}[H]
\centering
{
\begin{tabularx}{\textwidth}{cp{2.5cm}p{3cm}p{5cm}}
\hline
Symbol & Argument & Type & Definition\\
\hline
$P_t$ & $\mu,\sigma$ & Levy Process & {Mid-quote drift and volatility}\\
$N_t$ & $\lambda(t)$ & Poisson Process & {Spread jump intensity}\\
$\hat{S}_n$ & $(\rho_{ij})_{1\leq i,j \leq m}$ & {Discrete-time Markov Chain} & {Spread value jump process} \\
$S_t$ & $(r_{ij}(t))=\left(\lambda(t)\rho_{ij}\right)_{1\leq i,j \leq m}$ & {Continuous-time Markov Chain} & {Spread time and value jump process} \\
$N^a_t$ & $\lambda^a(Q^a_t,S_t)$ & Cox Process & Execution of limit sell \\
$N^b_t$ & $\lambda^b(Q^b_t,S_t)$ & Cox Process & Execution of limit buy \\
$X_t$ & - & - & Cash\\
$Y_t$ & - & - & Inventory \\
\hline
\end{tabularx}}
\caption{Table of State Variables}
\label{table:state_variables}
\end{table}

Four state variables are involved in this approach: the LOB mid-quote, the LOB bid-ask spread, the amount of cash, and the stock inventory of the maker.  Table~\ref{table:state_variables} lists related variables, and we shall provide a concise overview of each, more detailed exhibitions are referred to the Appendix.
\begin{itemize}

\item  The Exogenous LOB Mid-Quote $P_t$ 

The mid-quote $P_t$ of the stock price is assumed to follow a stochastic diffusion process,
\begin{align}
\dd{P_t} = \mu \dd{t} + \sigma \dd{W_t}, 0 \leq t \leq T,
\label{eq:pt}
\end{align}
where $W_t$ is a standard Brownian motion, $\mu$ and $\sigma$ are constants representing the drift and volatility, respectively.

\item  The Exogenous LOB Bid-Ask Spread $S_t$ 

A continuous-time finite state process $S_t \in \mathbb{S}$ is proposed to describe the dynamics of the bid-ask spread, where $\mathbb{S}=\delta\mathbb{I}_m$, $\mathbb{I}_m = \{1,\cdots,m\}$, and $m\in \mathbb{N}^{+}$ is a constant. Two independent processes $N_t$ and $\hat{S}_n$ are introduced to model the jump transition of $S_t$, where $N_t$ is a Poisson process for the cumulative count of random bid-ask spread jumps by time $t$, and $\hat{S}_n$ is a discrete-time Markov chain for the description of spread value after $n$th jumps. The spread process $S_t$ is characterized by $S_t = \hat{S}_{N_t}, t \geq 0$.  

\item Cash $X_t$ and Inventory $Y_t$ of the Maker

The cash $X_t$ represents the net monetary value of the market maker’s transactions, while the inventory $Y_t$ denotes the net amount of the traded asset that the market maker holds or owes. The dynamics of these state variables are influenced by the execution of limit and market orders fulfilled.

For a limit order as described in Equation~\eqref{eq: make2}, upon being filled, the cash $X_t$ and inventory $Y_t$ will be changed correspondingly as 
\begin{align}
dX_t &= -\pi^b(Q^b_t, P_{t^{-}}, S_{t^{-}}) L^b_t dN^b_t + \pi^a(Q^a_t, P_{t^{-}}, S_{t^{-}}) L^a_t dN^a_t \label{eq:xlimit} \\
dY_t &= L^b_t dN^b_t - L^a_t dN^a_t  , \label{eq:ylimit}
\end{align}
where $\pi^b(Q^b_t, P_{t^{-}}, S_{t^{-}})$ and $\pi^a(Q^a_t, P_{t^{-}}, S_{t^{-}})$, defined in Equation~\eqref{eq:pi_b},\eqref{eq:pi_a}, represent the bid and ask prices for the limit orders taking into the commission and stamp tax into account, respectively, and $L^b_t$ and $L^a_t$ denote the sizes of the limit buy and sell orders, respectively. The cash $X_t$ and inventory $Y_t$ are updated as the limit orders are filled, as indicated by the independent Cox processes $N^b_t$ and $N^a_t$, representing the cumulative execution count of limit buy and sell orders from the market maker by time $t$. The intensities of these processes depend on the bid/ask quotes $Q^b_t$ and $Q^a_t$, as well as the spread $S_t$, expressed as $\lambda^b(Q^b_t,S_t)$ and $\lambda^a(Q^a_t,S_t)$, respectively.

For market orders as described in Equation~\eqref{eq:takeAppex}, we assume that they are filled immediately upon submission at the best available quote in the LOB. Consequently, the cash $X_t$ and inventory $Y_t$ will be changed according to the market order executions as 
\begin{align}
  Y_{\tau_n} &= Y_{\tau^{-}_n} + \zeta_n,   \\
  X_{\tau_n} &= X_{\tau^{-}_n} - c(\zeta_n, P_{\tau_n}, S_{\tau_n}),
  \label{eq:xymarket}
\end{align}
where $c(\zeta_n, P_{\tau_n}, S_{\tau_n})$, as defined in Equation~\eqref{eq:c}, represents the amount of cash corresponding to this market order.

\end{itemize}

\subsubsection{Control Variables}
\begin{table}[H]
\centering
\begin{tabular}{cll}
\hline
Symbol & Range & Definition\\
\hline
$L^b_t$ & $[0,\bar{l}]$ & Size of limit buy order \\
$L^a_t$ & $[0,\bar{l}]$ & Size of limit sell order \\
$Q^b_t$ & $\mathbb{Q}^b$ & Quote of limit buy order \\
$Q^a_t$ & $\mathbb{Q}^a$ & Quote of limit sell order  \\ \hline
$\tau_n$ & $[0,T]$ & Submission time of $n^{th}$ market order\\
$\zeta_n$ & $[-\bar{e},\bar{e}]$ & Size of $n^{th}$ market order \\ \hline
\end{tabular}
\caption{Table of Control Variables}
\label{table:control_variables}
\end{table}

The market maker controls his or her the placement of limit and market orders.   Now we briefly introduce the control variables as listed in Table~\ref{table:control_variables}.
\begin{itemize}
\item The Limit Order Strategy $\alpha^{make}$ 

The limit order strategy is modeled as a continuous control process,
\begin{align}
\alpha_t^{make}=(Q^b_t,Q^a_t,L^b_t,L^a_t), \,\, 0 \leq t \leq T,
\label{eq: make}
\end{align}
where $L^b_t \in [0, \bar{l}]$ and $L_t^a \in [0, \bar{l}]$ represent the size of the buy and sell limit orders with maximum limit order size $\bar{l}$, and $Q^b_t \in \mathbb{Q}^{b}$ and $Q^a_t \in \mathbb{Q}^a$ represent the corresponding bid quote and ask quote, respectively. For example, Figure~\ref{fig:LOB} illustrates three-level bid quotes and ask quotes, in which $\mathbb{Q}^{b} = \{Bb-, Bb,Bb{+}\}$ and $\mathbb{Q}^a = \{Ba+, Ba,Ba{-}\}$. Here, $Ba$ and $Bb$ signify the best ask quote and best bid quote, respectively, while $\delta$ denotes one tick size.

\begin{figure}[h]
 	\centering
     \includegraphics[width=0.5\textwidth]{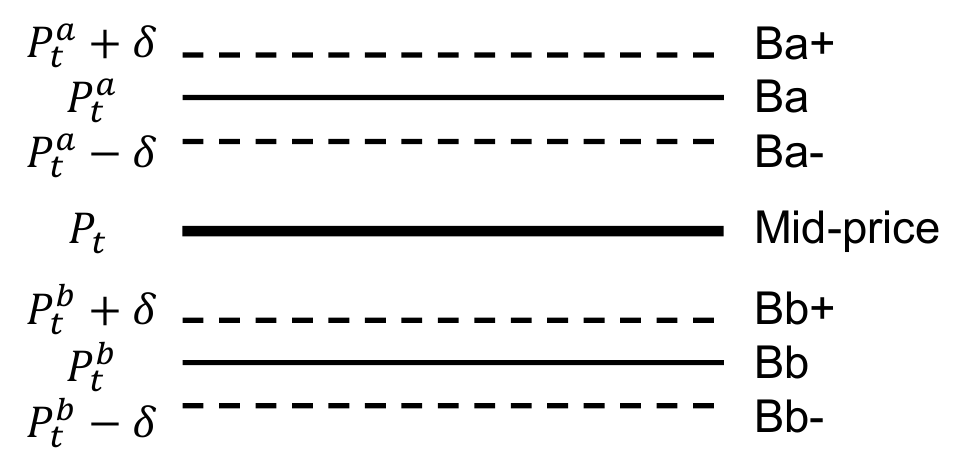}
 	\caption{An illustration of three-level bid and ask quote $Q_t^b, Q_t^a$. }
    \label{fig:LOB}
\end{figure}

\item The Market Order Strategy $\alpha^{take}$ 

The market order strategy is modeled as an impulse control,
\begin{equation}
    \alpha^{take}=(\tau_n,\zeta_n)_{n\geq 0},
    \label{eq:take}
\end{equation}
where $\tau_n \in [0,T]$ represents the time at which the market maker places the  $n^{th}$ market order, and $\zeta_n \in [-\bar{e}, \bar{e}]$ is a random variable representing the size of the $n^{th}$ market order. Here, $\bar{e}$ represents the maximum market order size. 

\end{itemize}

\subsubsection{Optimal Control Problem}
Upon identifying the state and control variables, the optimal trading strategy can be formulated by solving a stochastic optimal control problem. In this section, we present the formulation of this problem and its corresponding solution.

\begin{itemize}
\item  Problem Setup for the Trading Strategy

The optimal trading strategy aims to maximize the expectation of the terminal wealth while considering risk aversion towards inventory over the trading horizon. The optimal trading strategy, denoted by $\alpha = (\alpha^{make},\alpha^{take})$, is then determined by solving an optimization problem, given by
\begin{equation}
  \max_{\alpha} \mathbb{E}\left[U\left(L\left(X_{T}, Y_{T}, P_{T}, S_{T}\right)\right)-\gamma \int_{0}^{T} g\left(Y_{t}\right) \mathrm{d} t\right],  \label{eq:problem}
\end{equation}
where $U$ is a monotonically increasing reward function, $g$ is a non-negative convex function as inventory penalty, and $\gamma$ is a non-negative penalty constant  for the inventory risk aversion, $L(x, y, p, s)$ represents a liquidation function (e.g. the total amount of cash a trader would obtain if they were to immediately liquidate their entire position at the current market price).

\item Value Function

According to Equation~\eqref{eq:problem}, the value function for this optimal control problem is given by
\begin{align}
v(t,x,y,p,s)=\max_{\alpha} \mathbb{E}_{t, x,y,p,s}\left[U\left(L\left(X_{T},Y_{T},P_{T}, S_{T}\right)\right)-\gamma \int_{t}^{T} g\left(Y_{u}\right) \mathrm{d} u\right]
\label{eq:ocp}
\end{align}
where $\mathbb{E}_{t, x,y,p, s}$ denotes the expected value based on the underlying processes $(X,Y,P,S)$ with initial values $(X_{t^{-}},Y_{t^{-}},P_{t^{-}},S_{t^{-}})=(x,y,p,s)$. 
Since that the spread $s$ is discrete and takes values in $\mathbb{S} = \delta \mathbb{I}_m$, the value function for $s=i\delta$ can be expressed in a more convenient form as $v_i(t,x,y,p)=v(t,x,y,p,i\delta)$ for $i \in \mathbb{I}_m$.

\item Hamilton-Jacobi-Bellman (HJB) Equation

By incorporating limit and market order strategies as controls and employing Itô's lemma, the dynamic programming equation of Equation~\eqref{eq:ocp} can be expressed as
\begin{equation} \label{eq:hjb}
    \min \left[-\frac{\partial v_{i}}{\partial t}-\max_{(q, \ell) \in \mathbb{Q}^b\times \mathbb{Q}^a \times[0, \bar{\ell}]^{2}} \mathcal{L}^{q, \ell} v+\gamma g, v-\mathcal{M} v\right]_{i \in \mathbb{I}_m}=0,
\end{equation}
with the terminal condition $v(T, x, y, p, s)=U(L(x, y, p, s))$. Here, $\mathcal{L}^{q,l}$ and $\mathcal{M}$ denote the infinitesimal generators for limit and market order controls, respectively.

For limit order control, given $\alpha_t^{make}=(q,l) = (q^b,q^a, l^b,l^a)$, 
\begin{align}
\begin{split}
 \mathcal{L}^{q, \ell} v(t, x, y, p, s) &=\mathcal{L}_{P} v(t, x, y, p, s)  +{R}_S(t) v(t, x, y, p, s) \\
 &+A^b v(t,x,y,p,s) +A^a v(t,x,y,p,s),
\end{split}
\label{eq:lql}
\end{align}

where $\mathcal{L}_{P}$ represents the infinitesimal generator of the mid-quote process $P$, ${R}_S(t)$ represents the generator of the continuous-time Markov chain price process $S$, and $A^b, A^a$ represent the infinitesimal generators of the jump processes caused by the changes in cash and inventory when this limit order occurs. For the explicit expressions of these infinitesimal generators, please refer to Appendix~\ref{app:valuefunc}.

For market order control, the following impulse operator is considered
\begin{align}
\mathcal{M} v(t, x, y, p, s)=\max_{e \in[-\bar{e}, \bar{e}]} v\left(t, x-c(e, p, s), y+e, p, s\right) .
\label{eq:m}
\end{align}
\end{itemize}

\section{Numerical Scheme} \label{sec:num_tec}

The solution to the the optimal limit/market order controls of the maker is obtained by a numerical scheme to the HJB equation. Different from the classical finite difference method \cite{FORSYTH2007} to directly discretize the PDE, our numerical scheme is based on the definition of the infinitesimal generators of encompassed stochastic processes and their expectation forms. In this section we give the sketch of the numerical scheme. Despite leading to the identical numerical scheme as \cite{guilbaud2013optimal}, the explicit derivation in this paper is  generic and can be used for other similar problems.  

\subsection{Derivation of the Numerical Scheme}
With an equally-spaced partition over the time interval $[0, T]$,  time grid points are obtained as $\mathbb{T}_{n}=\left\{t_{k}=k h, k=0, \ldots, n\right\}$, where the time step size is $h=\frac{T}{n}$. Then, for the real-valued function $v_i(t,x,y,p) = v(t,x,y,p,s), t\in[0,T], x \in \mathbb{R}, y \in \mathbb{R}, p \in \mathbb{R}^+, s = i\delta \in \mathbb{S}$,

For the derivative of the value function $v_{i}$ with respect to time, the Euler scheme with the equally-spaced time step $h=\frac{T}{n}$ is written as, 
\begin{align} \label{eq:euler-time}
    \begin{split}
        \frac{\partial v_{i}}{\partial t} = & \left[v_{i}(t+h, x, y, p)- v_{i}(t, x, y, p)\right]/h  + o(h) \\
        \approx & \left[{v_{i}(t+h, x, y, p)-v_{i}(t, x, y, p)}\right]/{h}  \\
    \end{split}
\end{align}

By the definition of the infinitesimal generator,  $\mathcal{L}_p$ in Equation~\eqref{eq:lql} can be approximated by 
\begin{align} \label{eq:lp}
\begin{split}
\mathcal{L}_p v_{i}(t+h, x, y, p) = &
    \left\{\lim_{\hat{h} \to 0^+}\mathbb{E}\left[v_{i}\left(t+h, x, y, P_{t+\hat{h}}^{t, p}\right)\right]-v_{i}(t+h, x, y, p)\right\}/{\hat{h}} \\
    =&    \left\{\mathbb{E}\left[v_{i}\left(t+h, x, y, P_{t+\hat{h}}^{t, p}\right)\right]-v_{i}(t+h, x, y, p)\right\}/{\hat{h}} + o(\hat{h}) 
\end{split}
\end{align}
where $\hat{h}$  stands for a small time interval approaching to zero. Please note that $\hat{h}$ may be different from  time step size $h$ used for the time grid.  When choosing $\hat{h} = 4h$, Equation \eqref{eq:lp} becomes

\begin{align} \label{eq:lp-4h}
\mathcal{L}_p v_{i}(t+h, x, y, p) 
    \approx     \left\{\mathbb{E}\left[v_{i}\left(t+h, x, y, P_{t+4h}^{t, p}\right)\right]-v_{i}(t+h, x, y, p)\right\}/{(4h)},  
\end{align}
where the expectation is taken on random variable $P_{t+4h}^{t, p}$. 
Similarly, we can obtain the approximation of the other three generators as follows
\begin{align} \label{eq:rs-4h}
R_S(t)  v_{i}(t+h, x, y, p) 
    \approx     \left\{\mathbb{E}\left[v\left(t+h, x, y, p, S_{t+4 h}^{t, i \delta}\right)\right]-v_{i}(t+h, x, y, p)\right\}/{(4h)},  
\end{align}
{
\begin{align} \label{eq:ab-4h}
\begin{split}
A^b v_{i}(t+h, x, y, p) 
    \approx & \left\{\mathbb{E}\left[v _ { i } \left(t+h, x-\pi^{\mathrm{b}}\left(q^{\mathrm{b}}, p, i\delta\right) \ell^{\mathrm{b}} \Delta N_{4 h}^{i, q^{\mathrm{b}}}, y+\ell^{\mathrm{b}} \Delta N_{4 h}^{i, q^{\mathrm{b}}}, p\right)\right] \right. \\
    &  -v_{i}(t+h, x, y, p) \Bigl\}/{(4h)},  
\end{split}
\end{align}
\begin{align}  \label{eq:aa-4h}
\begin{split}
A^a v_{i}(t+h, x, y, p) 
    \approx & \Bigl\{\mathbb{E}\left[v _ { i } \left(t+h, x+\pi^{\mathrm{a}}\left(q^{\mathrm{a}}, p,i\delta\right) \ell^{\mathrm{a}} \Delta N_{4 h}^{i, q^{\mathrm{a}}}, y-\ell^{\mathrm{a}} \Delta N_{4 h}^{i, q^{\mathrm{a}}}, p\right)\right] \\
    & -v_{i}(t+h, x, y, p) \Bigl\} / (4h),  
\end{split}
\end{align}}
where function $\pi^a,\pi^b$ and $c$ are defined in Equation~\eqref{eq:pi_a},\eqref{eq:pi_b},\eqref{eq:c}.

By discretizing Equation~\eqref{eq:hjb} and putting together \eqref{eq:euler-time} to \eqref{eq:aa-4h}, we arrive at the operators 
\begin{align} \label{eq:Dh}
\mathcal{D}_{i}^{h}(t, x, y, p, v)=\max\left[\mathcal{T}_{i}^{h}(t, x, y, p, v), \mathcal{M}_{i}^{h}(t, x, y, p, v)\right],
\end{align}
where 
\begin{align} \label{eq:Th}
&\mathcal{T}_{i}^{h}(t, x, y, p, v) \notag \\
=&-h \gamma g(y)+\frac{1}{4}\left\{\mathbb{E}\left[v_{i}\left(t+h, x, y, P_{t+4 h}^{t, p}\right)\right]+\mathbb{E}\left[v\left(t+h, x, y, p, S_{t+4 h}^{t, i \delta}\right)\right]\right. \notag \\
&+\max_{\left(q^{\mathrm{b}}, \ell^{\mathrm{b}}\right) \in \mathbb{Q}^b \times[0, \bar{\ell}]} \mathbb{E}\left[v _ { i } \left(t+h, x-\pi^{\mathrm{b}}\left(q^{\mathrm{b}}, p, i\delta\right) \ell^{\mathrm{b}} \Delta N_{4 h}^{i, q^{\mathrm{b}}}, y+\ell^{\mathrm{b}} \Delta N_{4 h}^{i, q^{\mathrm{b}}}, p\right)\right] \notag \\
&+\max_{\left(q^{\mathrm{a}}, \ell^{\mathrm{a}}\right) \in \mathbb{Q}^a \times[0, \bar{\ell}]} \mathbb{E}\left[v _ { i } \left(t+h, x+\pi^{\mathrm{a}}\left(q^{\mathrm{a}}, p,i\delta\right) \ell^{\mathrm{a}} \Delta N_{4 h}^{i, q^{\mathrm{a}}} ,
y\left.-\ell^{\mathrm{a}} \Delta N_{4 h}^{i, q^{\mathrm{a}}}, p\right)\right]\right\} ,
\end{align}
and
\begin{align} \label{eq:Mh}
\mathcal{M}_{i}^{h}(t, x, y, p, v)=\max_{e \in[-\bar{e}, \bar{e}]} v_{i}\left(t +h, x-c(e, p, i\delta), y+e, p\right).    
\end{align}
In this formula, $P^{t,p}$ represents the mid-quote Markov process that starts at time $t$ with price $p$, while $S^{t,i\delta}$ denotes the spread process that begins at time $t$ with a spread of $i\delta$. Furthermore, $\Delta N^{i,q^b}_h$ indicates the increment of a Poisson process with a rate of $\lambda(q^b, i\delta)$ over the interval $[t, t + h]$. Similarly, the same applies to $\Delta N^{i,q^a}_h$.

Subsequently, the Euler-scheme discretized solution with step size $h$, represented by $v^h = (v_i^h)_{i \in \mathbb{I}_m}$, can be determined by solving backward in time using the following formulas
\begin{align}\label{eq:v_h}
\begin{split}
v_{i}^{h}\left(t_{n}, x, y, p\right)&=U\left(L_{i}(x, y, p)\right), \\
 v_{i}^{h}\left(t_{k}, x, y, p\right)&=\mathcal{D}_{i}^{h}\left(t_{k}, x, y, p, v^{h}\right), k=n-1, \cdots, 0 .
\end{split}
\end{align}

\subsection{Convergence of the Numerical Scheme to HJB Equation}
We prove the convergence of the scheme below. For $\mathcal{T}_{i}^{h}(t, x, y, p, v)$, from the definition of the infinitesimal generator and \cite{john2021calculating}, we have
\begin{align}
\mathbb{E}\left[v_{i}\left(t+h, x, y, P_{t+4 h}^{t, p}\right)\right]&=v_{i}(t+h, x, y, p)+4 h \mathcal{L}_p v_{i}+o(h) \notag\\
\mathbb{E}\left[v_i\left(t+h, x, y, \rho, S_{t+4 h}^{t, i \delta}\right)\right]&=v_{i}(t+h, x, y, \rho)+4 h R_S(t) v_{i}+o(h) \notag \\
\mathbb{E}\left[v_{i}\left(t+h, x-\pi_{i}^{b} l^{b} \Delta N_{4 h}^{i, q^{b}}, y+l^{b} \Delta N_{4 h}^{i, q^{b}}, p\right)\right]&=v_{i}(t+h, x, y, p)+4 h A^{b} v_{i}+o\left(h\right) \notag \\
\mathbb{E}\left[v_{i}\left(t+h, x+\pi_{i}^{a} l^{a} \Delta N_{4 h}^{i, q^{a}}, y-l^{a} \Delta N_{4 h}^{i, q^{a}}, p\right)\right]&=v_{i}(t+h, x, y, p)+4 h A^{a} v_{i}+o\left(h\right). \label{eq:4e}
\end{align}

From Equation~\eqref{eq:euler-time}, \eqref{eq:Th}, \eqref{eq:4e}, we have
\begin{align}
\lim _{h \rightarrow 0} \frac{v_{i}(t, x, y, p)-\mathcal{T}_{i}^{h}(t, x, y, p, v)}{h} =-\frac{\partial v_{i}}{\partial t}-\max_{(q, \ell) \in \mathbb{Q}^b \times \mathbb{Q}^a \times[0, \bar{\ell}]^{2}} \mathcal{L}^{q, \ell} v_{i}+\gamma g. 
\end{align}

For $\mathcal{M}_{i}^{h}(t, x, y, p, v)$, it is the direct finite difference form of $\mathcal{M}_i$, and thus
\begin{align}
\lim _{h \rightarrow 0} [v_{i}(t, x, y, p) -\mathcal{M}_{i}^{h}(t, x, y, p, v) ]= v_{i}-\mathcal{M} v_{i}.
\end{align}
Combining these two equations above, we have
\begin{align}
\begin{split}
&\lim _{h \rightarrow 0} \min \left[\frac{v_{i}(t, x, y, p)-\mathcal{T}_{i}^{h}(t, x, y, p, v)}{h}, v_{i}(t, x, y, p) -\mathcal{M}_{i}^{h}(t, x, y, p, v)\right] \\
    &=\min \left[-\frac{\partial v_{i}}{\partial t}-\max_{(q, \ell) \in \mathbb{Q}^b \times \mathbb{Q}^a \times[0, \bar{\ell}]^{2}} \mathcal{L}^{q, \ell} v_{i}+\gamma g, v_{i}-\mathcal{M} v_{i}\right],    
\end{split}
\end{align}
which indicates that when $h \rightarrow 0$, Equation~\eqref{eq:v_h} converges to the solution of the HJB equation.

\section{Numerical Experiments} \label{sec:result}

In this section, we present the results of our numerical experiments on market making in the Chinese stock market. We begin by discussing the parameter estimation process used to calibrate our model to actual market data. Subsequently, we compare the performance of our baseline strategy under various market conditions, including the impact of volatility, stamp duty, and price drift. Through our analysis and discussions, we provide valuable insights into market making strategies in the Chinese stock market and their implications for market makers and policymakers.

\subsection{Parameter Estimation}
\label{sec:est}
In this section, we focus on the process of estimating the key parameters using historical stock data from a historical period $[0, T_p]$, as well as present some results obtained from the analysis.

\subsubsection{Estimation of Spread and Mid-quote Processes}
The spread process $S_t = \hat{S}_{N_t}$ contains two components as discussed in Section~\ref{sec:state}: the discrete-time Markov chain $\hat{S}_n$ and the jump process $N_t$. We estimate the transition matrix $P=(\rho_{ij})$ and intensity $\lambda$ associated with these two components from stock data.

Since the actual value of spread process $S_t$ can be observed from the historical period $[0, T_p]$, we can determine the jump times of the spread process as 
$$\theta_0=0, \theta_{n+1} = \inf\{t>\theta_n : S_t \neq S_{t-}\}, \forall n \geq 1, $$
where $S_{t^-} = \lim_{p \rightarrow t^-} S_p$. Consequently, we can deduce the actual values of $N_t$ and $\hat{S}_n$ from their definitions by
$$\begin{aligned}
N_t &= \#\{j: 0<\theta_j \leq t\}, t \geq 0, \\
\hat{S}_n &= S_{\theta_n}, n \geq 0,
\end{aligned}$$
where $\#\{\cdot\}$ denote the size of the set. Knowing the actual values of these two component processes, we can then estimate their parameters using maximum likelihood estimation (MLE).
For the transition matrix $(\rho_{ij})_{1\leq i,j \leq m}$ of $\hat{S}_n$, a consistent estimator of its element $\rho_{ij}$ can be derived by
\begin{align}
\hat{\rho}_{i j}=\frac{\sum_{n=1}^{N_{T_p}} 1_{\left\{\left(\hat{S}_{n}, \hat{S}_{n-1}\right)=(j \delta, i \delta)\right\}}}{\sum_{n=1}^{N_{T_p}} 1_{\left\{\hat{S}_{n-1}=i \delta\right\}}}.
\label{eq:rho_est}
\end{align}
Meanwhile, a consistent estimator for the intensity $\lambda$ of the jump process $N_t$ is 
\begin{align}
    \hat{\lambda}=\frac{N_{T}}{T}.
    \label{eq:lba_est}
\end{align}

For the estimation of the mid-quote process, assuming $R$ snapshots of $P_t$ are observed at $\{t_i = i\Delta t : i=0, \cdots, R, \Delta t = \frac{T}{R} \}$, consistent estimators for $\mu$ and $\sigma^2$ are given by
\begin{align}
    \hat{\mu} &= \frac{P_T -P_0}{T}, \label{eq:mu_est}\\
    \hat{\sigma}^2 &= \frac{\sum_{n=1}^{R} (P_{t_n}-P_{t_{n-1}}- \hat{\mu})^2}{T}.
    \label{eq:sigma_est}
\end{align}

\subsubsection{Estimation of Limit Order Execution Processes}
Whether the limit order placed by our strategies get fully executed is modeled by two independent Cox processes $N_t^b$ and $N_t^a$, with intensity modeled by $\lambda^b(Q^b_t,S_t)$, $\lambda^a(Q^a_t,S_t)$.

Assuming the trader can observe the execution processes $N_t^b,N_t^a$ in real-time, which represent the number of limit orders at bid quote $Q^b_t$ and ask quote $Q^a_t$ are fully executed , respectively, then the observed trade data can be described as a five-tuple:
$$\left(N_{t}^{\mathrm{a}}, N_{t}^{\mathrm{b}}, Q_{t}^{\mathrm{a}}, Q_{t}^{\mathrm{b}}, S_{t}\right) \in \mathbb{R}^{+} \times \mathbb{R}^{+} \times \mathbb{Q}^a \times \mathbb{Q}^b \times \mathbb{S}, \quad t \in[0, T].$$
Assuming $N_t^a$ and $N_t^b$ are independent, since their intensity estimating procedures are the same, here we describe the estimation of $\lambda^b$. For $\lambda^b(q^b,s), q^b \in \mathcal{Q}^b, s\in \mathbb{S}$, the ratio of the volume of buy orders traded in the system while in state $(q^b,s)$ to the total time spent in that state is a consistent estimator.
The following point process is defined to describe the part of process $N_t^b$ belonging to $(q^b,s)$, that is, the part where the trade quote of limit buy orders is $q^b$ and the spread is $s$: 
$$N_{t}^{\mathrm{b}, q^{\mathrm{b}}, s}=\int_{0}^{t} \mathbbm{1}_{\left\{Q_{u}^{\mathrm{b}}=q, S_{u-}=s \right\}} \mathrm{d} N_{u}^{\mathrm{b}}, \quad t \geq 0.$$
Similarly, the time that the system state stays in $(q^b,s)$ can be defined as:
$$\mathcal{T}_{t}^{\mathrm{b}, q^{\mathrm{b}}, s}=\int_{0}^{t} \mathbbm{1}_{\left\{Q_{u}^{\mathrm{b}}=q, S_{u-}=s\right\}} \mathrm{d} u.$$
Therefore, a consistent estimator for $\lambda^b(q^b,s)$ is:
$$ \hat{\lambda}^{\mathrm{b}}\left(q^{\mathrm{b}},s\right)=\frac{N_{T}^{\mathrm{b}, q^{\mathrm{b}}, s}}{\mathcal{T}_{T}^{\mathrm{b}, q^{\mathrm{b}}, s}}.$$

\begin{table}[htb]
\centering
\begin{tabular}{c|ccc}
\textbf{Spread} & $\bm \delta$ & $\bm{2\delta}$ & $\bm{3\delta}$ \\ \hline
$\bm \delta$ & 0 & 0.9404 & 0.0596 \\
$\bm{2\delta}$ & 0.9557 & 0 & 0.0443 \\
$\bm{3\delta}$ & 0.4074 & 0.5925 & 0 \\
\end{tabular}
\caption{Estimation of the transition matrix of $\hat{S}_n$ for the stock Ping An Bank Co. Ltd. (000001.SZ) on August 29, 2019.}
\label{table:transition}
\end{table}

\begin{figure}[h]
 	\centering
 	
 	\subfigure[Buy limit order]{\includegraphics[width=0.4\textwidth]{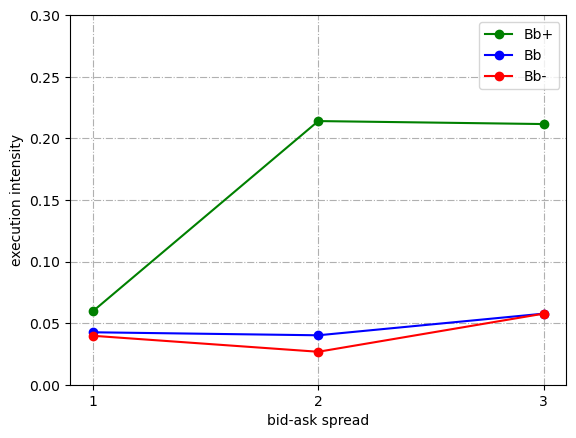}} 
	\subfigure[Sell limit order]{\includegraphics[width=0.4\textwidth]{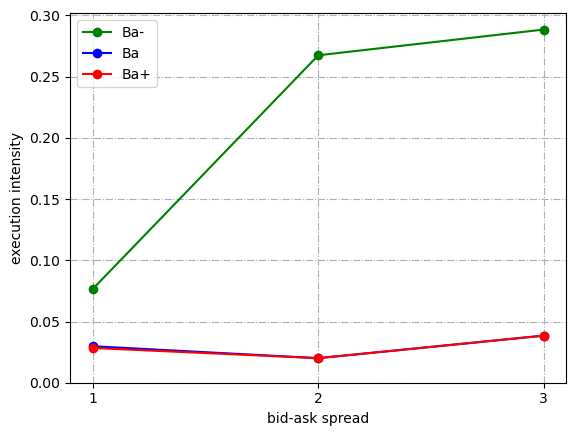}}
	
 	\caption{Estimation of the intensity of limit order execution processes for the stock Ping An Bank Co. Ltd. (000001.SZ) on August 29, 2019. }
  \label{fig:execution}
\end{figure}

\subsection{Baseline strategy}
To comprehensively evaluate the effectiveness of the strategy under varying market regulations and parameters within the Chinese stock market, we initially present a baseline strategy $\alpha^{\text{baseline}}$ by considering an ideal market.  For a liquidity provider, it is ideal to have both commission fee rate $\varepsilon$ and stamp duty fee rate $\rho$ set to $0\%$. We also assume that the designated market makers have pre-arrangements in place that make short selling of stocks feasible during market making.

We calibrate the parameters in the model according to Section~\ref{sec:est} using stock Ping An Bank Co. Ltd. (000001.SZ) on August 29, 2019. For the mid-quote process $P_t$ in Equation~\eqref{eq:pt}, we assume $\mu=0$ to consider that traders lack information about the stock price trend in the worst-case scenario. From the target stock data, we set $P_0 = 14$ Yuan; the volatility in seconds $\sigma=0.005$; the calibrated transition matrix $\rho_{ij}$ is presented in Table~\ref{table:transition} and the intensity $\lambda=1$ for the spread process $S_n$; and the calibrated execution intensities for the execution processes  $N_t^a$  and  $N_t^b$ are illustrated in Figure~\ref{fig:execution}.  
We consider solving the strategy for the trading period of duration $T=300$ seconds. The maximum allowable sizes for both limit and market orders are set to $\bar{l} = \bar{e} = 100$, while the inventory is constrained within the interval of $[y_{min}, y_{max}] = [-500, 500]$. To optimize Equation~\eqref{eq:problem}, we adopt the CARA utility function $U(x,y,p,s) = -\exp(-\eta (x-c(-y, p, s)))$ with $\eta = 0.5$, $c(y,p,s)$ in Equation~\eqref{eq:c}, and set $\gamma = 0$ as the optimization target. This utility function is equivalent to that of \cite{avellaneda2008high}, where the inventory risk is managed in an implicit way. Since the variables and functions are all determined in our optimal control problem, our control strategy $\alpha^{\text{baseline}}$ can be derived by solving the discrete numerical scheme in Equation~\eqref{eq:v_h}. Here we adopt $h = 0.3$ seconds as the discrete time step size.

In the backtesting process, we employ the Monte Carlo method to simulate the stock price movement  and apply the corresponding strategy. We generate a total of $100,000$ paths to represent various scenarios for the stock. For every individual path, we initialize both cash and inventory to be zero at $t=0$, and then proceed to simulate and monitor their evolution in accordance with the strategy implementation. 

\begin{figure}[h]
 	\centering
 	
    \subfigure[Mid-quote $P_t$]{\includegraphics[width=0.45\textwidth]{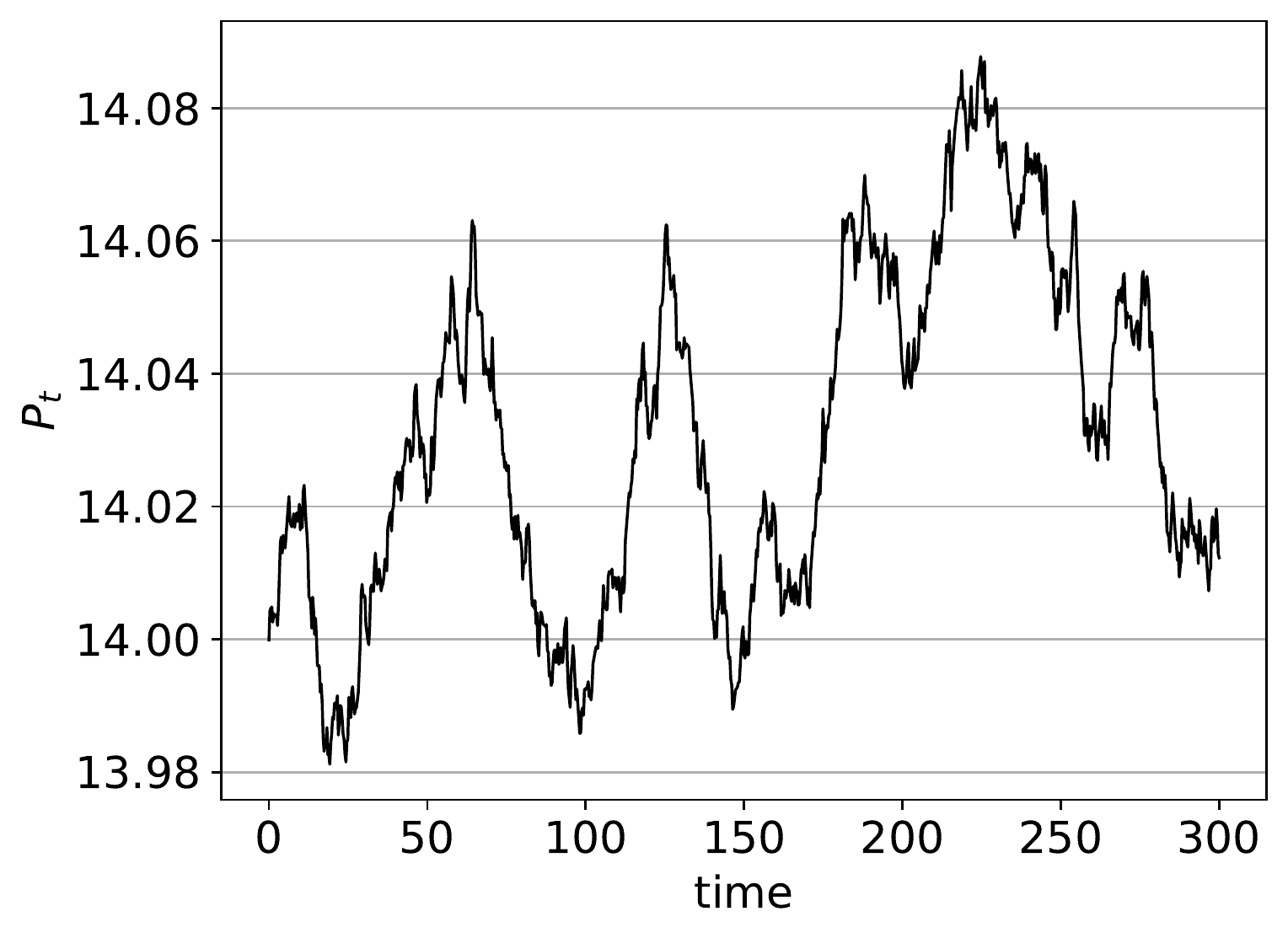}} \hspace{0.2cm}
    \subfigure[Spread $S_t$]{\includegraphics[width=0.42\textwidth]{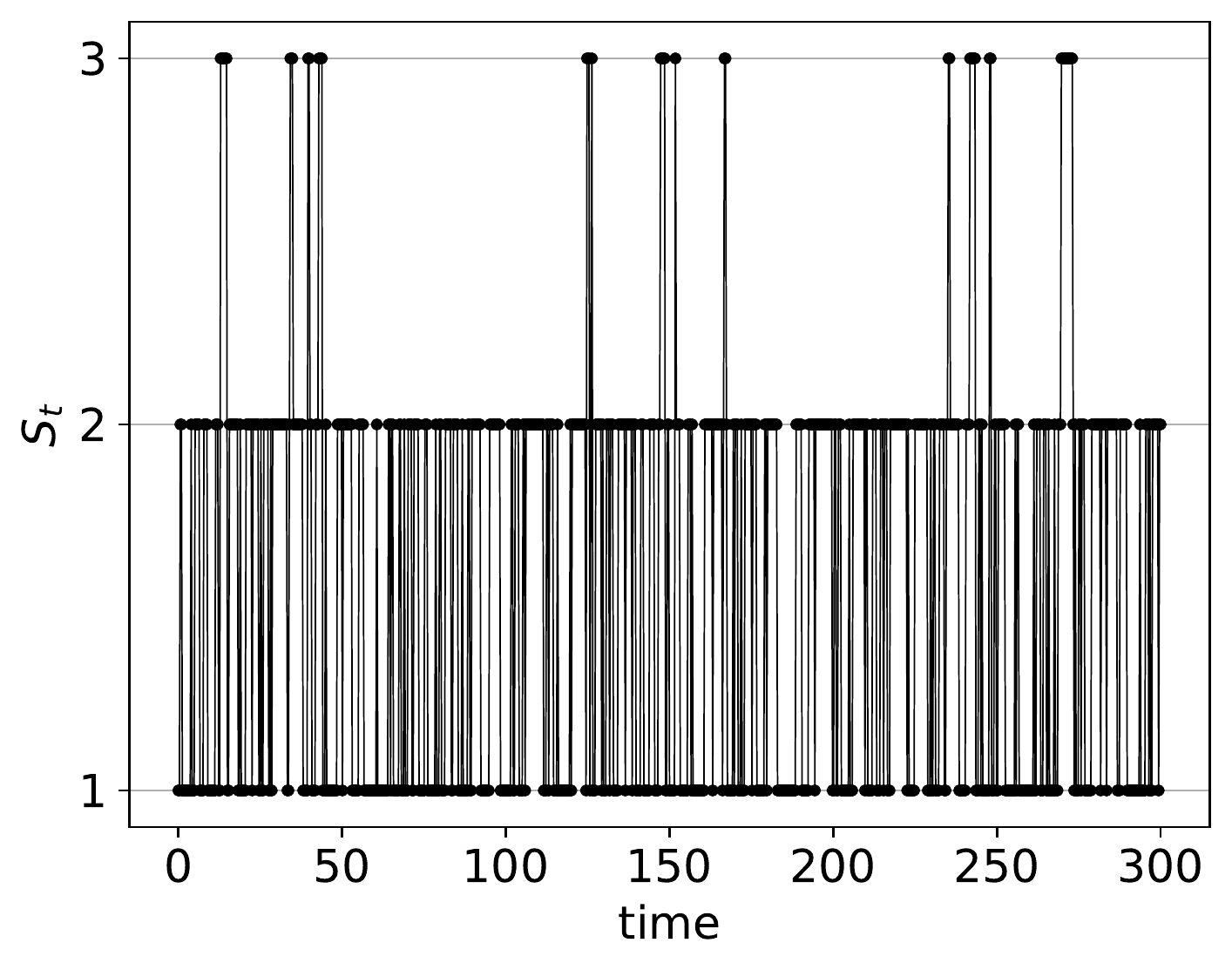}} 
    \subfigure[Absolute cumulative trading stocks $Q_t$]{\includegraphics[width=0.45\textwidth]{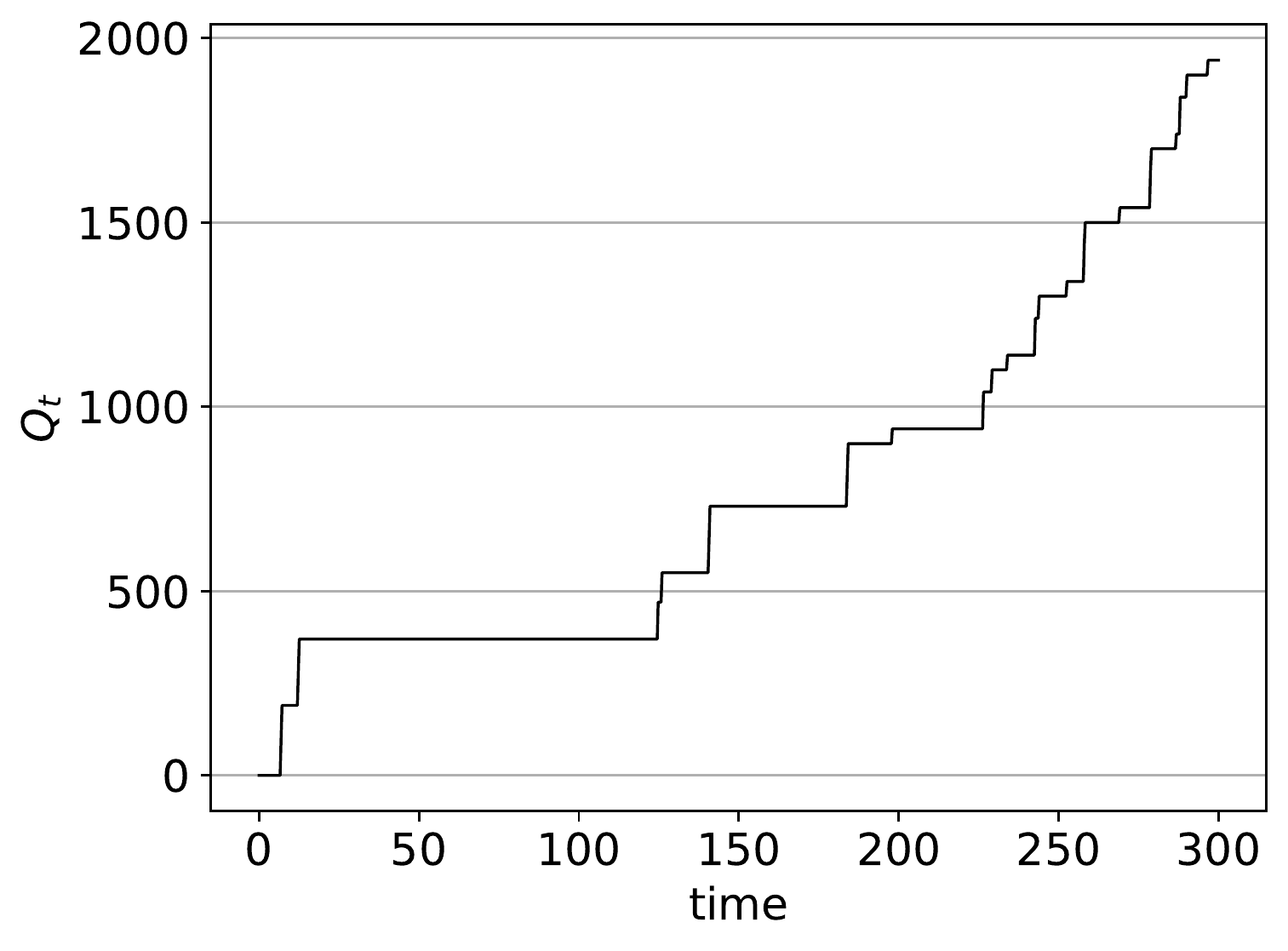}} 
    \subfigure[Cumulative wealth $U_t$]{\includegraphics[width=0.45\textwidth]{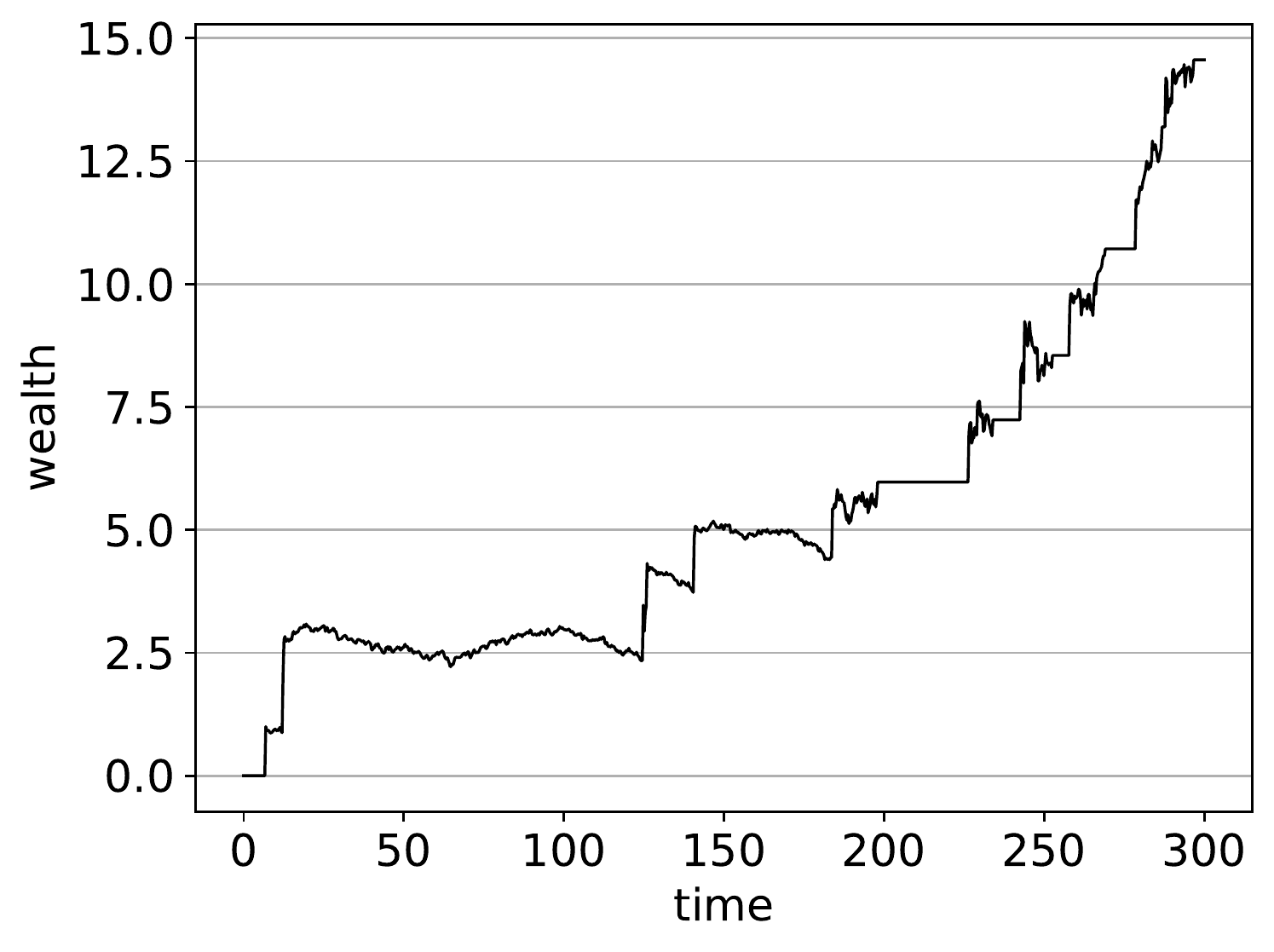}}

 	\caption{Evolution of Key Variables in a Single Backtest Path.}
    \label{fig:baseline_one_path}
\end{figure}

In Figure~\ref{fig:baseline_one_path}, we illustrate the evolution of key variables along one path during the backtest, including the mid-quote process $P_t$, the spread process $S_t$, the absolute cumulative trading stocks $Q_t$, and cumulative wealth $U_t$. Specifically, the absolute cumulative trading stocks $Q_t$ represent the absolute number of shares that the trader has successfully bought or sold through limit or market orders within the time interval $[0,t]$. The cumulative wealth $U_t$ accounts for the total wealth at time $t$, encompassing both cash from spread profit and stock value, as expressed by $U_t = X_t - c(Y_t,P_t,S_t)$ according to Equation~\eqref{eq: liquidation}. This demonstrates that with time going on, the market maker uses the optimal strategy to profit from the bid-ask spread.

\begin{figure}[htb]
    \centering
    \subfigure[Empirical distribution of the profit $X_T$]{\includegraphics[width=0.48\textwidth]{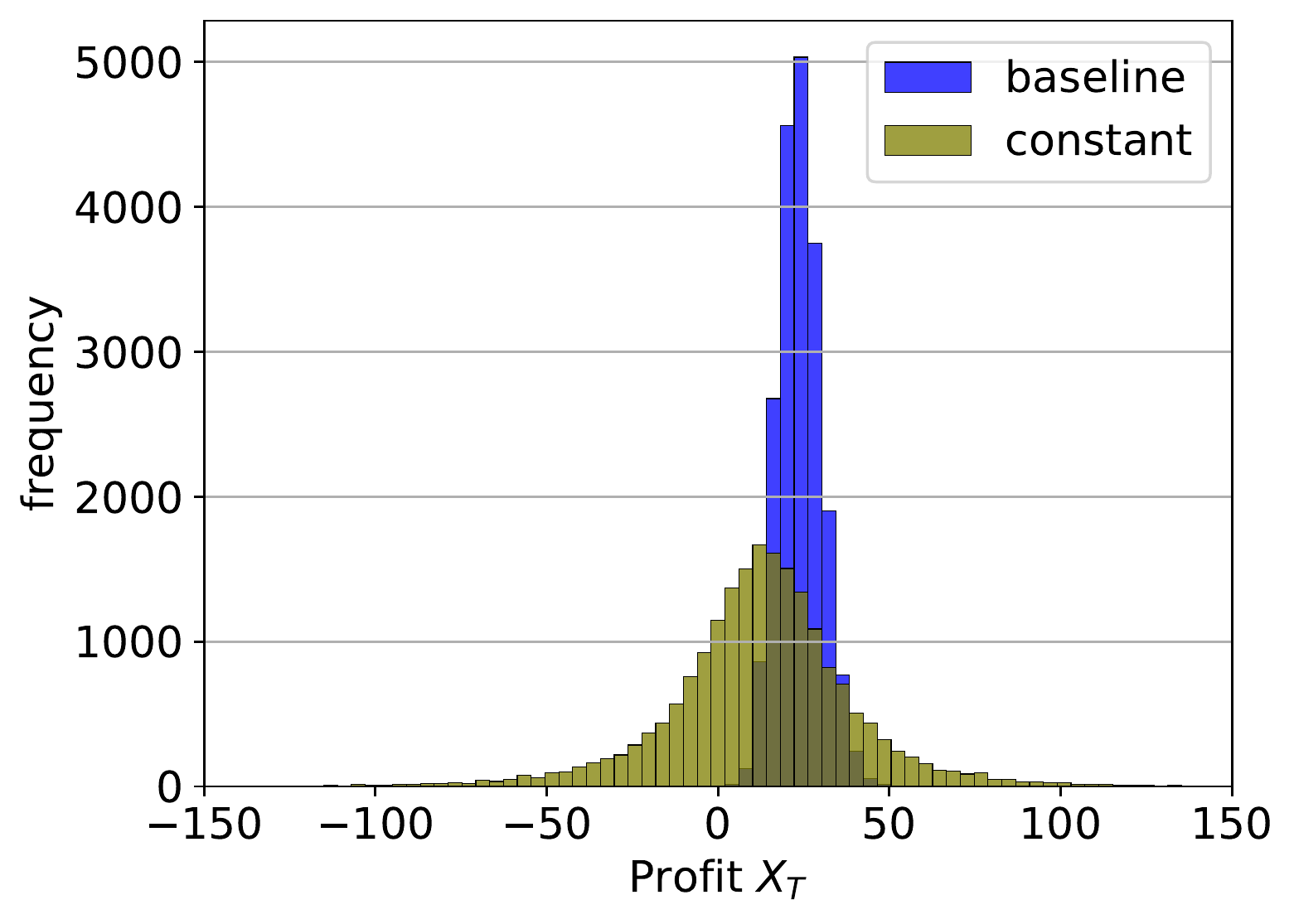}} 
    \subfigure[Empirical distribution of Mean($|Y_t|$)]{\includegraphics[width=0.48\textwidth]{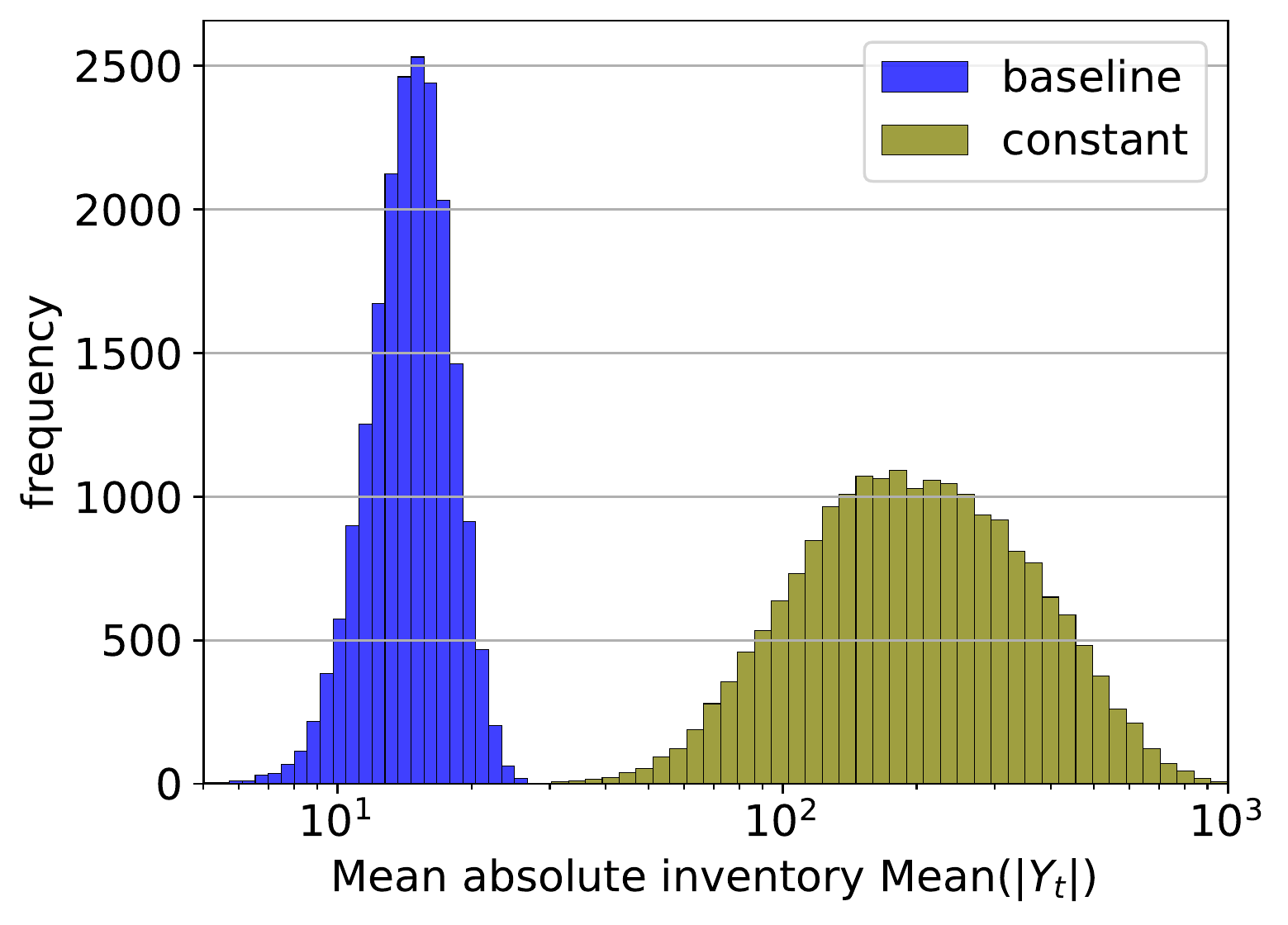}} 
    \caption{Comparison of the empirical distribution of profit $X_T$ and the mean absolute inventory Mean($|Y_t|$) between the Baseline strategy and Constant strategy. }
    \label{fig: emp}
\end{figure}

\begin{table}[thb]
\centering
\begin{tabular}{|l|l|l|l|}
\hline Quantity & Definition & $\alpha^{\text{baseline}}$ & $\alpha^{\text {constant}}$ \\
\hline Information Ratio & $m\left(X_T\right) / \sigma\left(X_T\right)$ & $3.850$ & $0.456$ \\
\hline Profit per trade & $m\left(X_T\right) / m\left(Q_T\right)$ & $0.0074$ & $0.0054$ \\
Risk per trade & $\sigma\left(X_T\right) / m\left(Q_T\right)$ & $0.0019$ & $0.0119$ \\
\hline Mean profit & $m\left(X_T\right)$ & $23.99$ & $12.88$ \\
Standard deviation of perf & $\sigma\left(X_T\right)$ & $6.23$ & $28.28$ \\
Skew of perf & $\operatorname{skew}\left(X_T\right)$ & $0.266$ & $-0.347$ \\
Kurtosis of perf & $\operatorname{kurt}\left(X_T\right)$ & $3.06$ & $7.37$ \\
\hline Mean total executed volume & $m\left(Q_T\right)$ & $3245.60$ & $2374.72$ \\
Mean at market volume & $m\left(Q^{\text {market}}_T\right)$ & $1163.49$ & $356.87$ \\
Ratio market over total exec & $m\left(Q^{\text {market}}_T\right) / m\left(Q_T\right)$ & $0.358$ & $0.150$ \\
\hline
\end{tabular}
\caption{Comparisons of evaluation metrics between baseline and constant strategies.}
\label{table: cmp}
\end{table}

To establish a benchmark for the performance of the Baseline strategy $\alpha^{\text{baseline}}$, we introduce the Constant strategy $\alpha^{\text{constant}}$ (e.g., a zero-intelligence agent) for comparison purposes. The Constant strategy involves placing buy and sell limit orders at the best bid and ask prices, respectively, with a maximum limit order size of $\bar{l}$ at each time step. No market orders are placed except at the final time $T$. Specifically, the limit order part of this strategy is defined as $\alpha^{\text{constant}, make}_t = (Bb, Ba, \bar{l}, \bar{l})$, in contrast to Equation~\eqref{eq: make}.

We compare the Baseline and Constant strategies using multiple metrics, as shown in Table~\ref{table: cmp}. The Baseline strategy demonstrates two primary advantages over the Constant strategy: higher profit per trade and lower risk per trade. The mean profit, profit per trade, and skewness of the Baseline strategy are notably higher than those of the Constant strategy, indicating that the Baseline strategy generates higher returns on average and exhibits a more positively-skewed distribution of profits. Conversely, the standard deviation, risk per trade, and kurtosis of the Baseline strategy are significantly lower than those of the Constant strategy, suggesting that the Baseline strategy incurs lower volatility in its performance. The lower kurtosis also indicates a more normally distributed profit for the Baseline strategy, implying fewer extreme outcomes compared to the Constant strategy. 

Overall, these advantages lead to a substantial improvement in the information ratio, increasing from 0.456 to 3.850, emphasizing the superior risk-adjusted performance of the Baseline strategy. Moreover,  Figure~\ref{fig: emp}(a) displays the empirical distributions of the profit for both strategies, further illustrating the enhanced performance of the Baseline strategy. The distribution of the Baseline strategy is more concentrated with a higher peak, suggesting a greater probability of achieving profits within a specific range and a more consistent performance compared to the Constant strategy.  Figure~\ref{fig: emp}(b) shows the empirical distributions of path-wise  mean absolute inventory Mean($|Y_t|$), computed as the average of the absolute inventory $|Y_t|$ for each path. By leveraging market orders, the Baseline strategy maintains a lower absolute inventory, as seen in Figure~\ref{fig: emp}(b), while preserving a similar level of limit orders as the Constant strategy, as indicated in Table~\ref{table: cmp}. This ensures consistent profits from limit orders while more effectively managing inventory risk.

\subsection{Volatility Impact}
Volatility measures the uncertainty of an asset price. The efficacy of the strategy may be affected by the degree in the volatility of the asset. Therefore, it is crucial to understand the impact of volatility on the performance of the strategy. In this section, we shall investigate the sensitivity of the performance with respect to the volatility. Furthermore, we shall assess how the levels of volatility may impact the strategy's inventory.

\subsubsection{The Impact of Volatility on Inventory}
\begin{figure}[h]
 	\centering
 	
 	\subfigure[Mean of Absolute Inventory]{\includegraphics[width=0.45\textwidth]{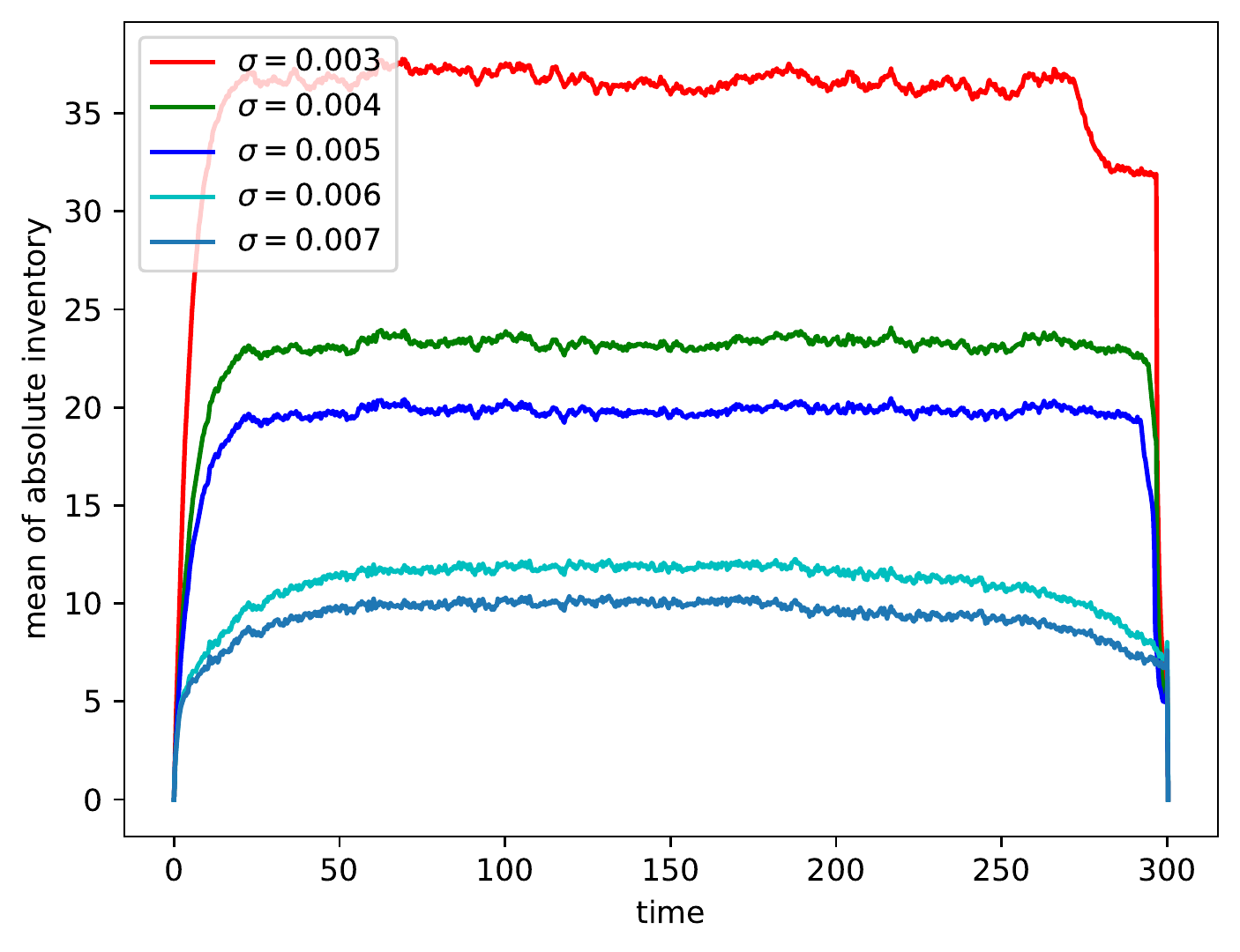}} 
    \subfigure[Std of Absolute Inventory]{\includegraphics[width=0.45\textwidth]{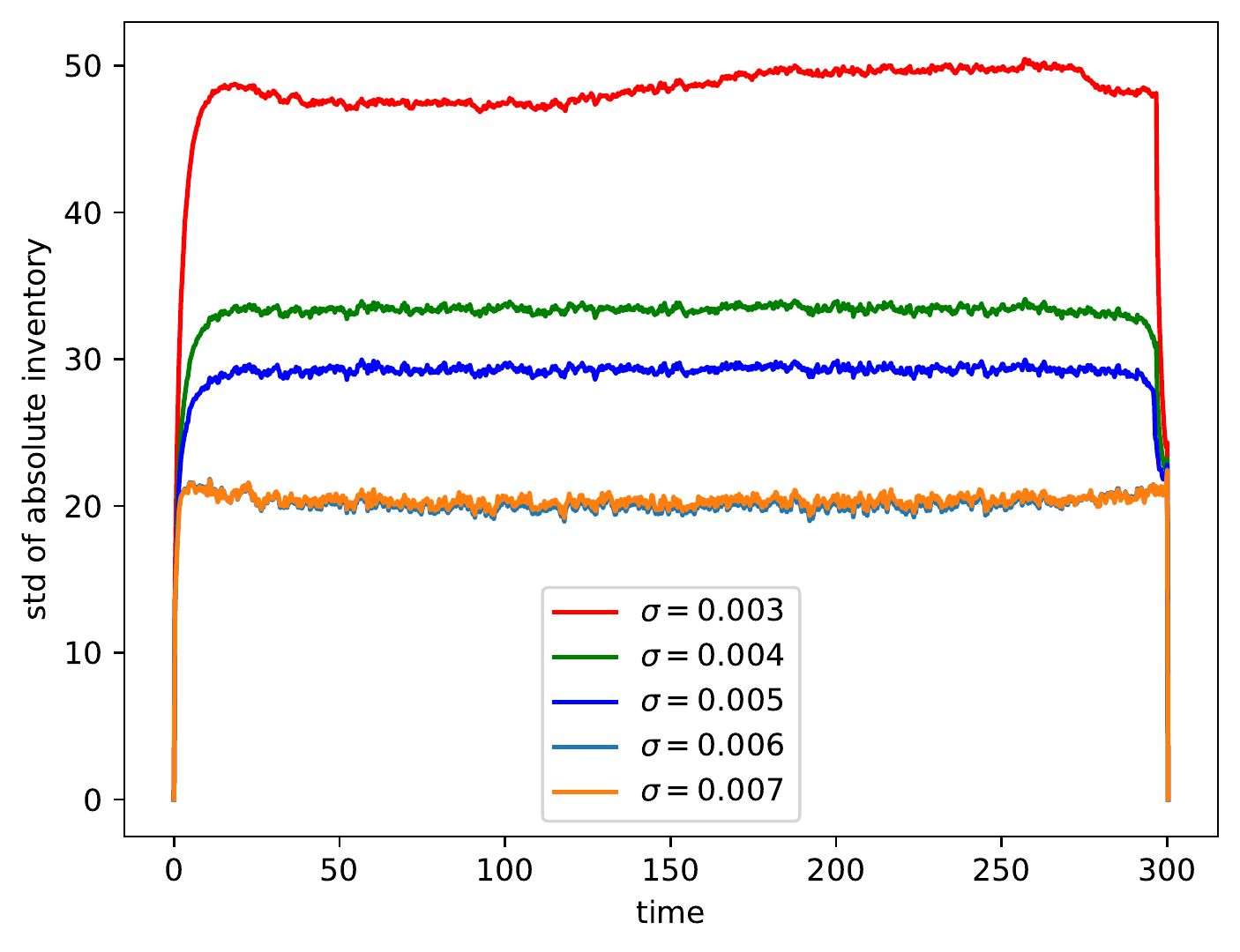}} 
 	\caption{Absolute inventory vs. volatility.}
    \label{fig:inventory}
\end{figure}

It is advisable for the strategy to adapt its inventory to varying levels of volatility to minimize inventory risk. To explore the impact of volatility on inventory, we replace the baseline strategy's volatility parameter, $\sigma$, with five distinct levels and calculate the corresponding strategy solutions. We then backtest these solutions and plot the curves of the mean and standard variance of the absolute inventory, $|Y_t|$, against time in Figure~\ref{fig:inventory}. 
The results reveal a negative correlation between absolute inventory and volatility. Specifically, as volatility rises, the probability of significant price fluctuations increases considerably. Consequently, to address the heightened price uncertainty, the strategy adjusts its inventory more frequently to mitigate the risks associated with such movements, resulting in a relatively lower absolute inventory. In contrast, when volatility is low, the strategy can afford to maintain its inventory for extended period of time since price movements are expected to be minor. In this scenario, the absolute inventory tends to be comparatively higher.

\subsubsection{Sensitivity w.r.t. Volatility} 

To assess the strategy's robustness, we subject the baseline strategy (derived for $\sigma=0.005$) to backtesting under various levels of perturbed volatility and evaluate its performance. Table~\ref{table:cali}  reveals that the mean profit remains relatively stable across a range of volatility levels. Conversely, the standard deviation of profit increases with higher values of $\sigma$, leading to a decline in the information ratio. This outcome suggests that the strategy can maintain its performance when volatility is overestimated. 

However, underestimating volatility negatively impacts the information ratio. A potential reason for the minor influence of volatility changes on the mean profit is that the model does not account for the correlation between volatility and intensity of execution processes. It is reasonable to assume a relationship between these factors: higher volatility levels might result in more intense trading activities, while lower levels could lead to reduced trading. 

\begin{table}[htb]  
\centering  
\begin{tabular}{|c|c|c|c|c|c|c|}  
\hline  
$\sigma$ & \multicolumn{2}{c|}{Mean of profit} & \multicolumn{2}{c|}{Std of profit} & \multicolumn{2}{c|}{Information Ratio} \\  
\cline{2-7}  
& Value & Change & Value & Change & Value & Change \\  
\hline  
0.003 & 23.951 & $-0.18\%$ & 5.843 & $-6.31\%$ & 4.099 & $+6.47\%$ \\  
\hline  
0.004 & 23.947 & $-0.20\%$ & 6.094 & $-2.20\%$ & 3.929 & $+2.05\%$ \\  
\hline  
0.005 & 23.994 & -- & 6.237 & -- & 3.850 & -- \\  
\hline  
0.006 & 23.963 & $-0.13\%$ & 6.458 & $+3.54\%$ & 3.710 & $-3.64\%$ \\  
\hline  
0.007 & 23.954 & $-0.16\%$ & 6.750 & $+8.23\%$ & 3.549 & $-7.82\%$ \\  
\hline  
\end{tabular}  
\caption{Sensitivity Analysis of Profit Metrics to Volatility Levels}  
\label{table:cali}  
\end{table}

\subsection{The Impact of Stamp Duty }
\begin{figure}[hbtp]
 	\centering
 	
 	\subfigure[Mean of profit $X_T$]{\includegraphics[width=0.45\textwidth]{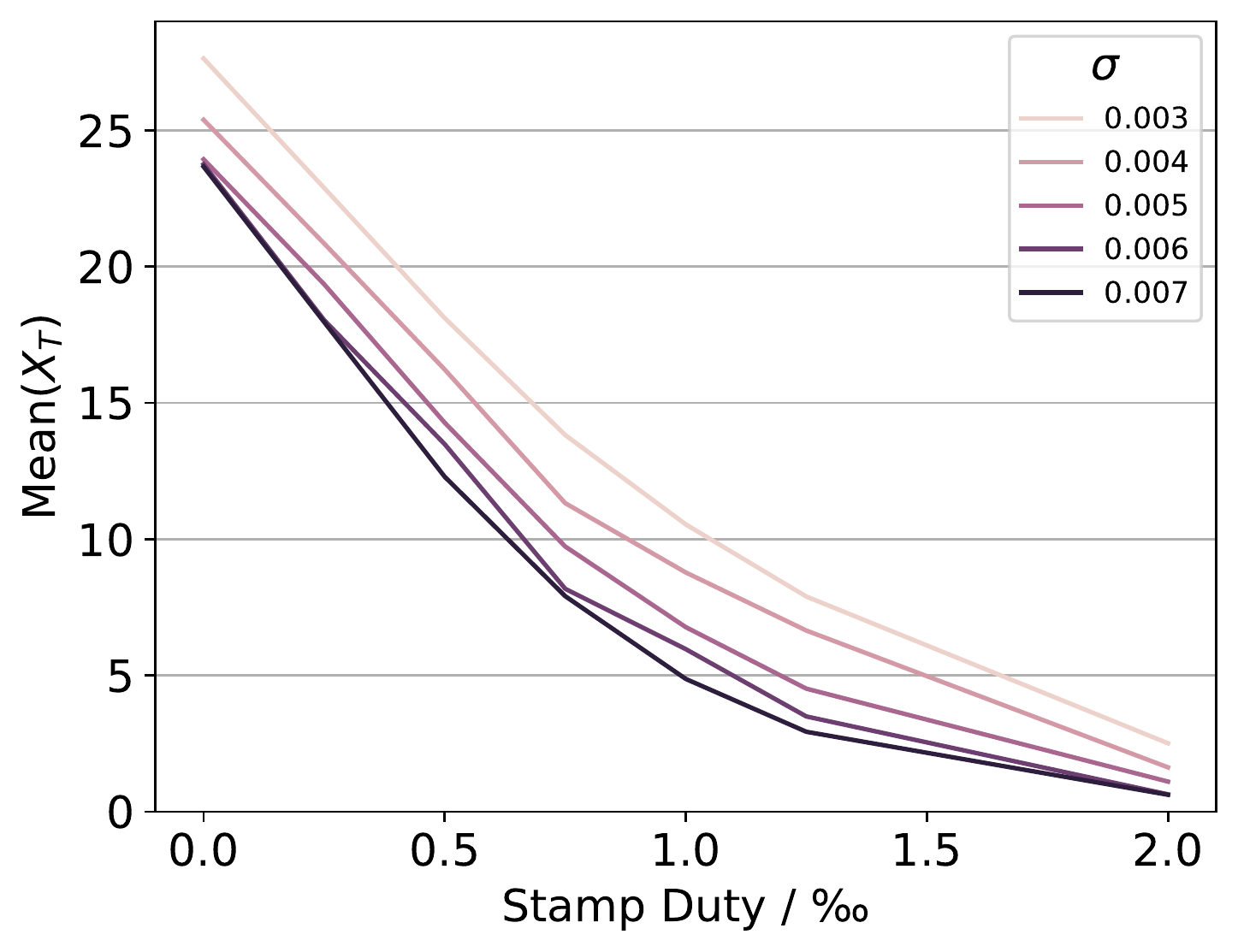}} 
    \subfigure[Std of profit $X_T$]{\includegraphics[width=0.45\textwidth]{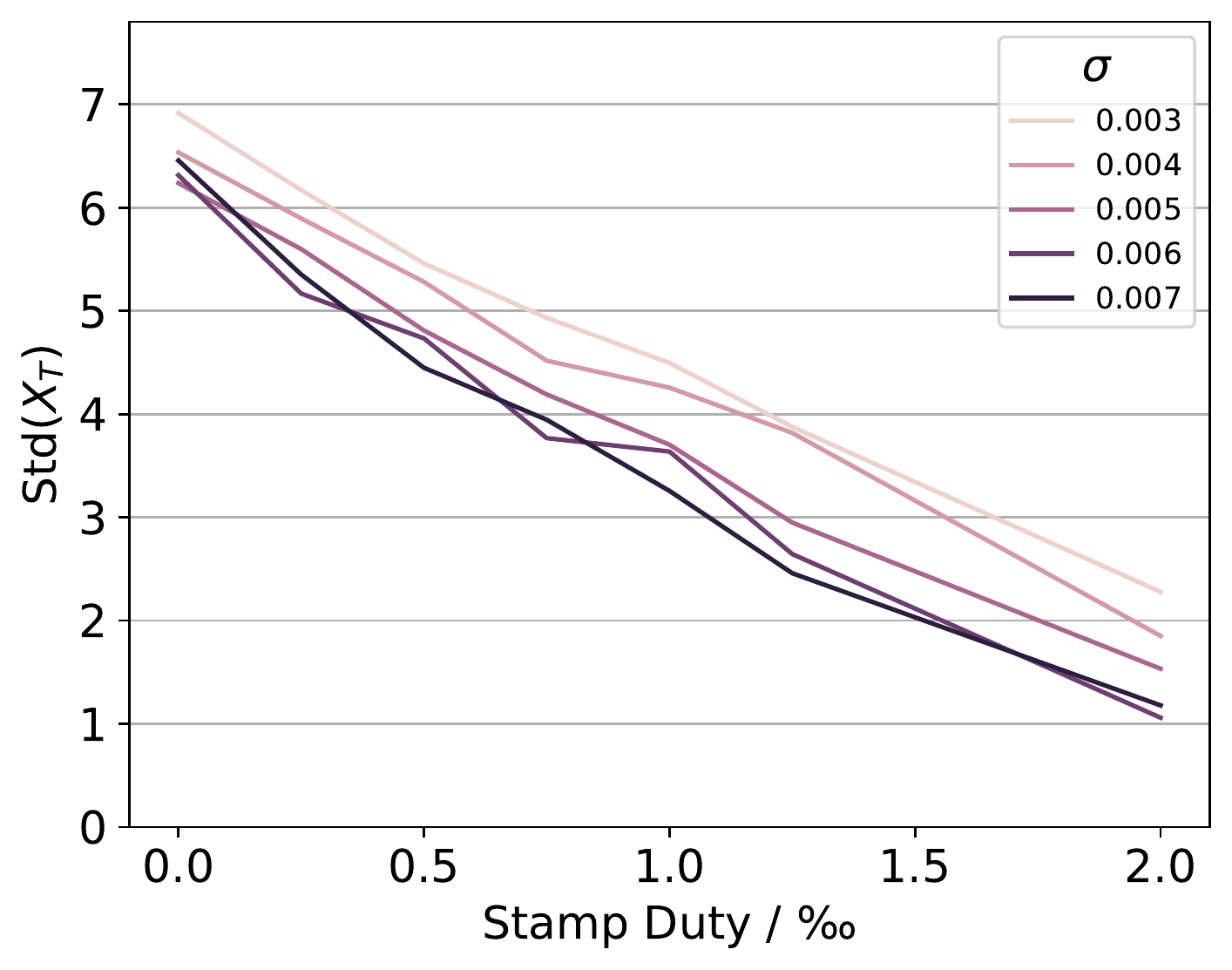}} 
    \subfigure[Mean of total executed volume $Q_T$]{\includegraphics[width=0.45\textwidth]{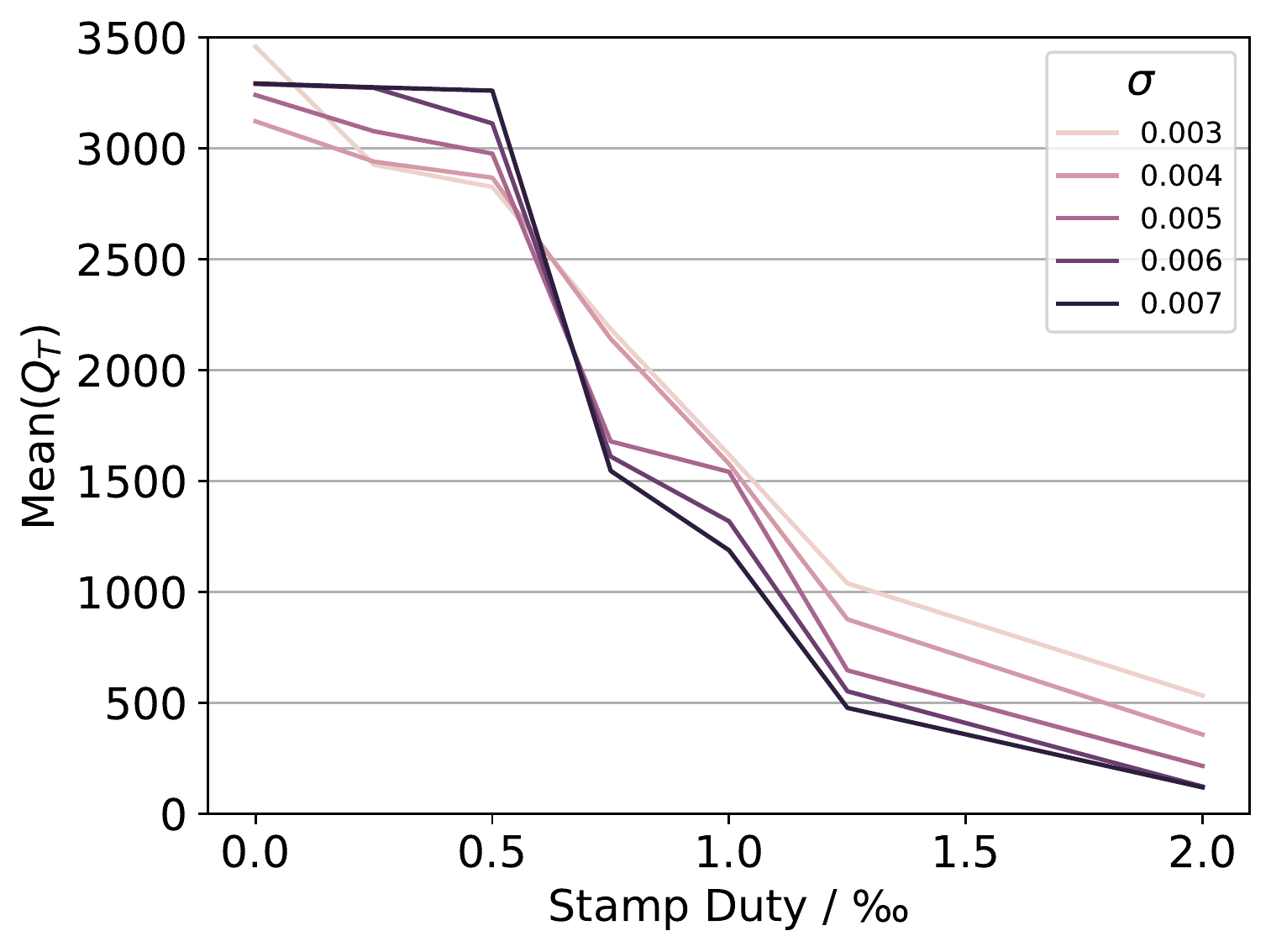}} 
    \subfigure[Std of total executed volume $Q_T$]{\includegraphics[width=0.45\textwidth]{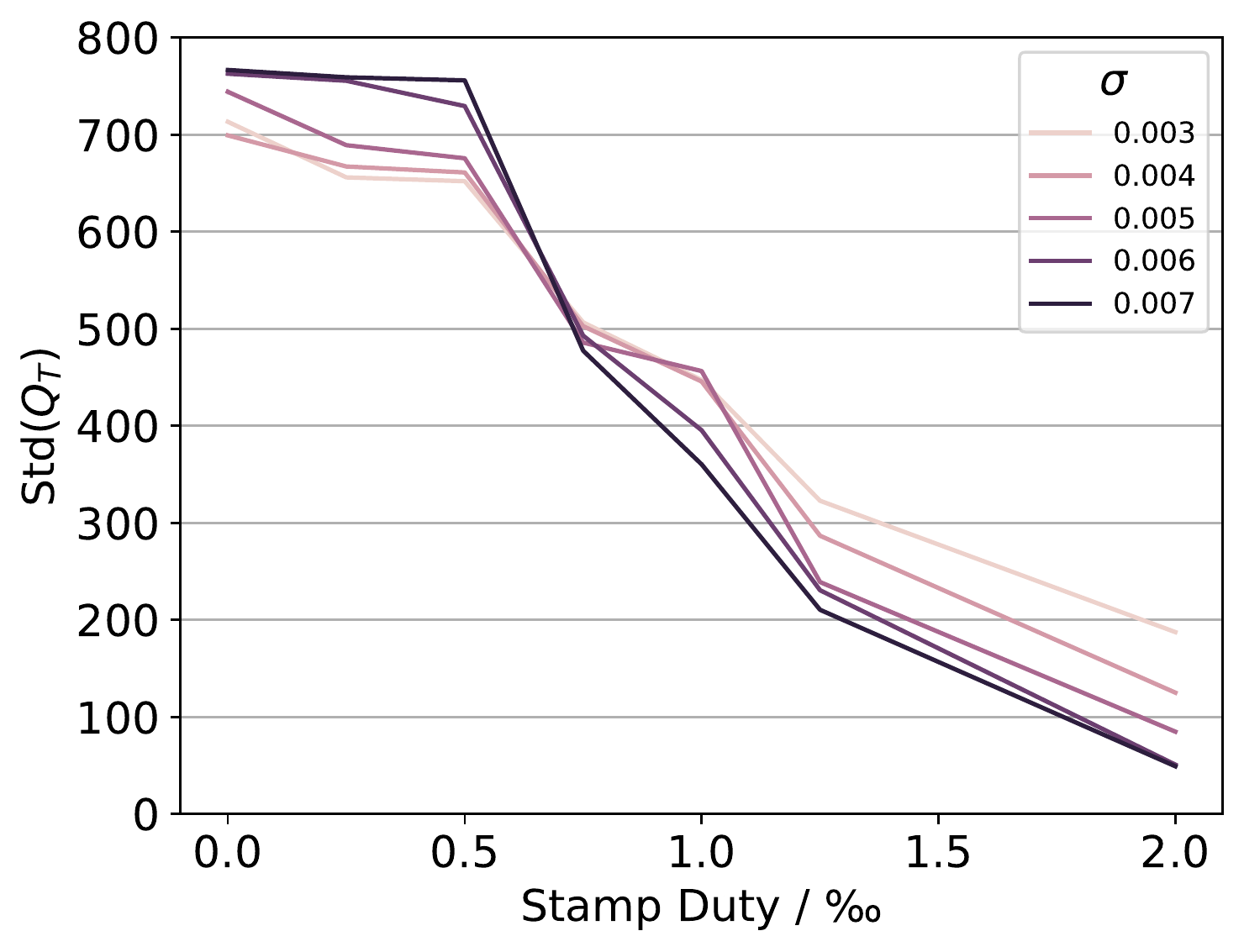}} 
    \subfigure[Mean of total tax]{\includegraphics[width=0.45\textwidth]{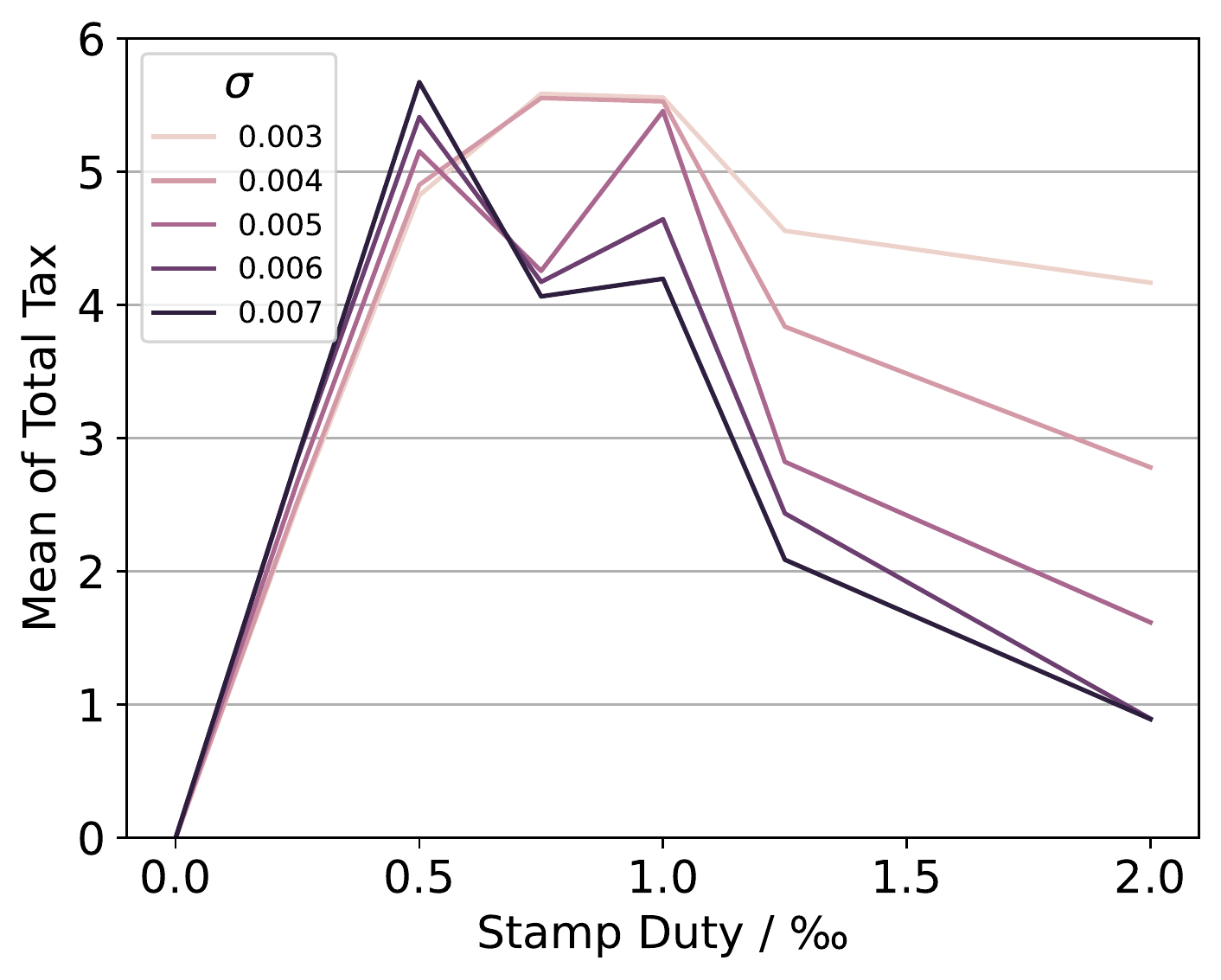}} 
    \subfigure[Std of total tax]{\includegraphics[width=0.45\textwidth]{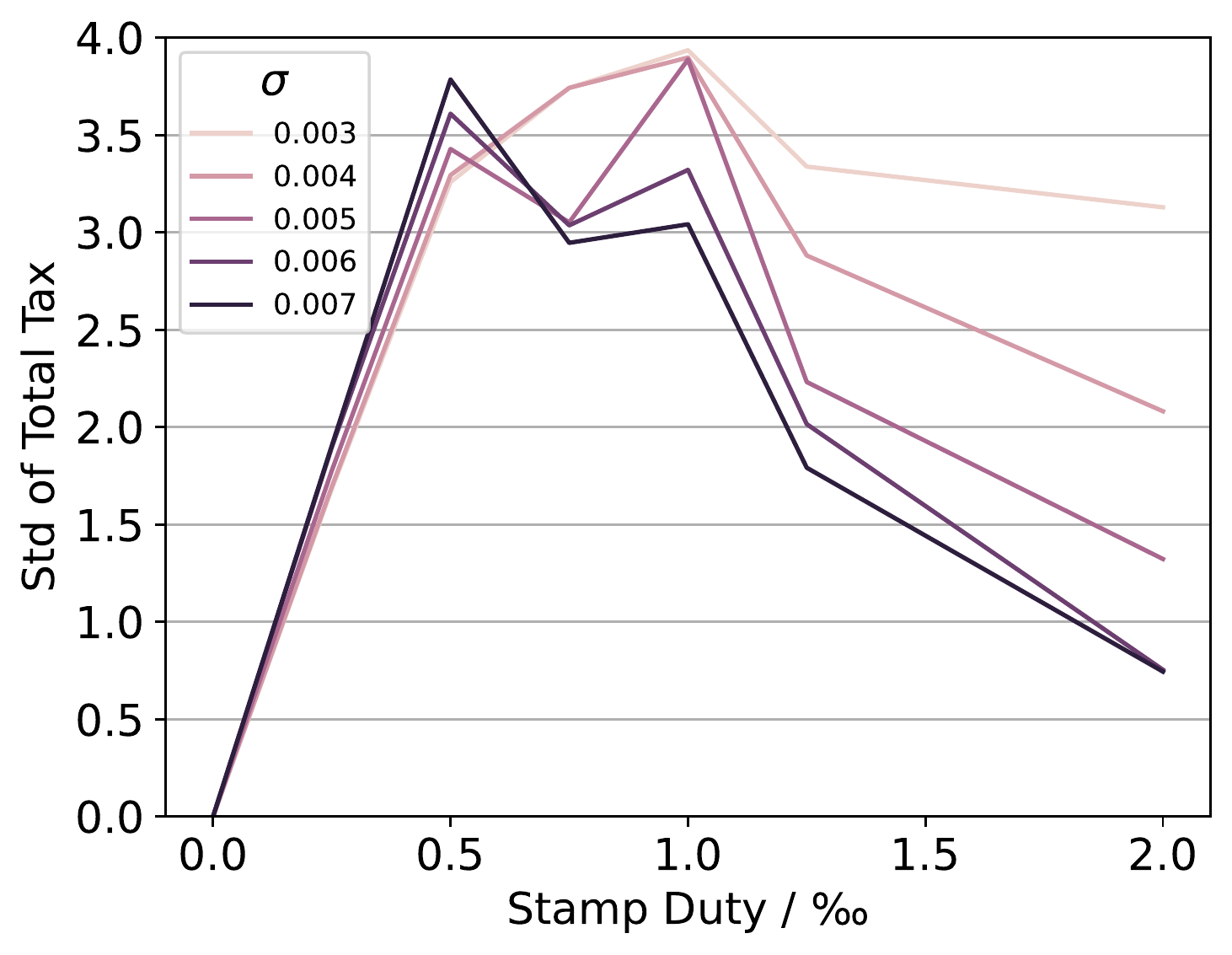}} 
 	\caption{profit $X_T$, total executed volume $Q_T$ and total tax vs. stamp duty.}
    \label{fig:stampduty}
\end{figure}

Stamp duty, a tax on financial transactions such as selling securities, can impact both market makers and government revenue. While it serves as a revenue source for governments, it can also affect market liquidity. In this section, we examine the impact of stamp duty on the Chinese market's liquidity, considering the implications for both market makers and government revenue.

We analyze the strategy under varying stamp duty rates, from 0‰ to 2‰, and different levels of volatility, from 0.003 to 0.007, for each stamp duty rate level. We backtest these strategies and keep track of the mean and standard deviation of profit, total executed volume, and total tax paid. The results are depicted in Figure~\ref{fig:stampduty}.

Figures~\ref{fig:stampduty}(a) and (b) illustrate that the profits of the maker decline almost exponentially as stamp duty rates increase under all volatility levels. This implies that stamp duty represents a crucial determinant of the profits for high-frequency market makers. 
Figures~\ref{fig:stampduty}(c) and (d) reveal a significant reduction in the total executed volume with higher stamp duty rates, suggesting that market liquidity is adversely impacted. Regarding government revenue, Figures~\ref{fig:stampduty}(e) and (f) reveal that the total tax paid by the strategy (in other words, collected by the government) initially increases and subsequently decreases as the stamp duty rate rises. Specifically in our experiment, when the stamp duty rate falls within the 0.5‰ to 1‰ range, the total tax paid barely fluctuates. From these results, we notice that with the increase of the stamp tax rate, both the market maker profit and the market liquidity (indicated by the total execution volume of the market making strategy) decrease rapidly, and the collected tax amount decreases with the tax rate in the high tax rate range. Therefore, an proper stamp duty rate should consider the trade-off between tax revenue, market liquidity, and market maker profits, ensuring a balance that benefits both market participants and government objectives.

\subsection{Drift Impact}

In this section, we explore the impact of drift on market making in the Chinese stock market. Drift, denoted by $\mu$, refers to the expected change in stock price over time and is defined in Equation~\eqref{eq:pt}. Accurate forecast of price drift is essential for the success of market making strategies, as it allows market makers to adjust their order strategies effectively and manage inventory and adverse selection risks. Therefore, witht the capacity of the current model framework, we aim to investigate the effect of variations in drift on the strategy and the its resulting performance. 

\begin{figure}[H]
 	\centering
 	
 	\subfigure[Drift $\mu=-0.001$]{\includegraphics[width=0.32\textwidth]{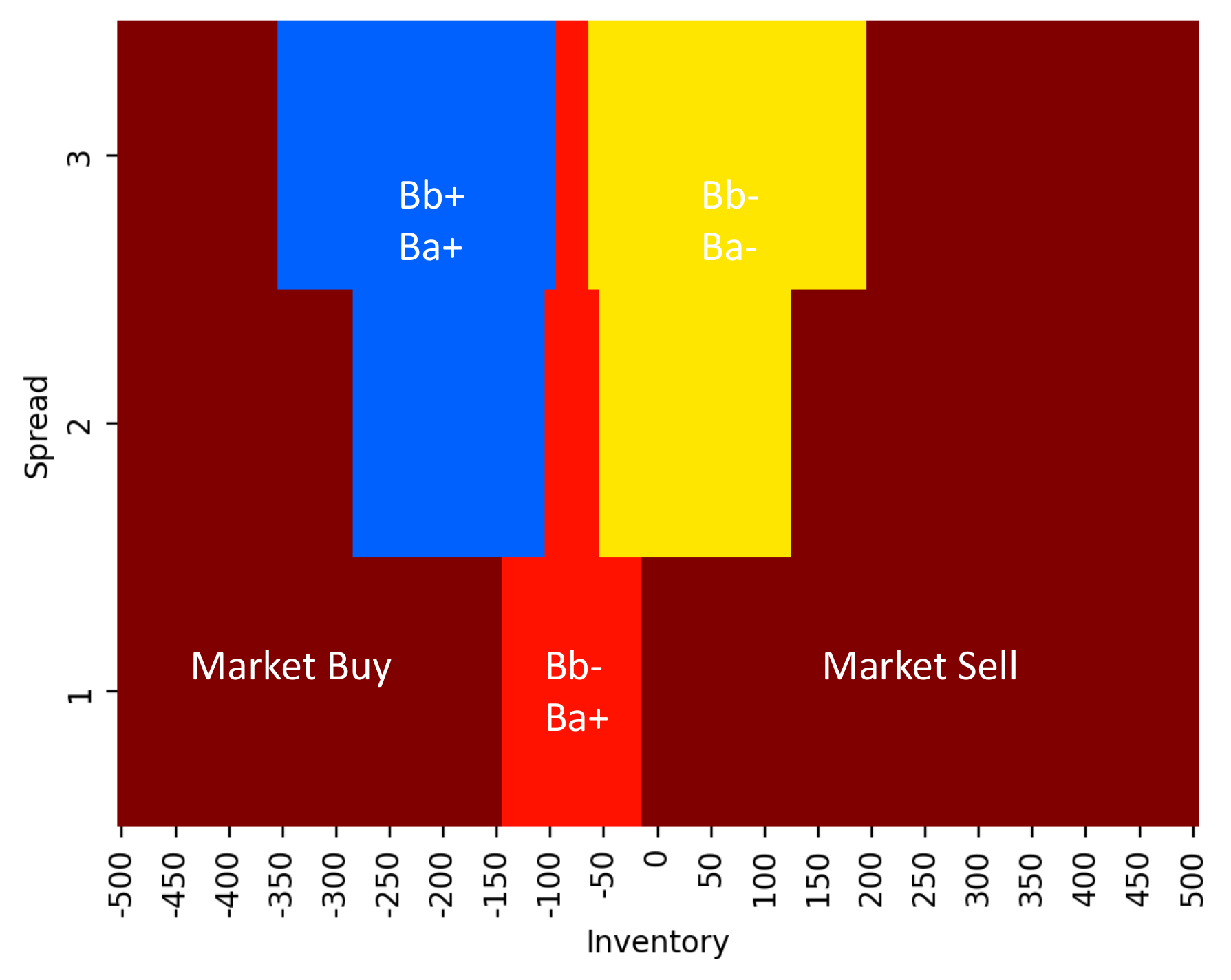}} 
    \subfigure[Drift $\mu=0$]{\includegraphics[width=0.32\textwidth]{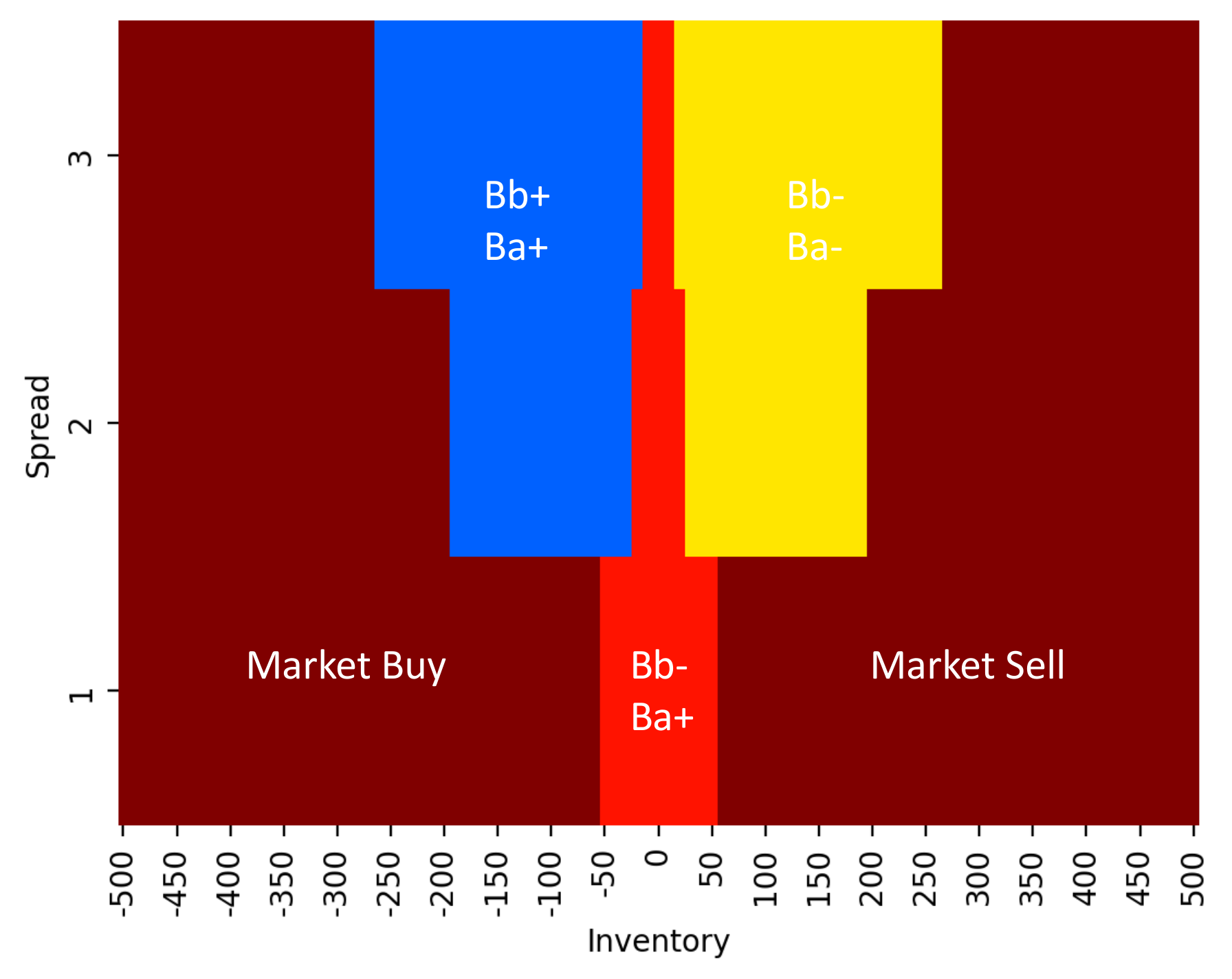}}
    \subfigure[Drift $\mu=+0.001$]{\includegraphics[width=0.32\textwidth]{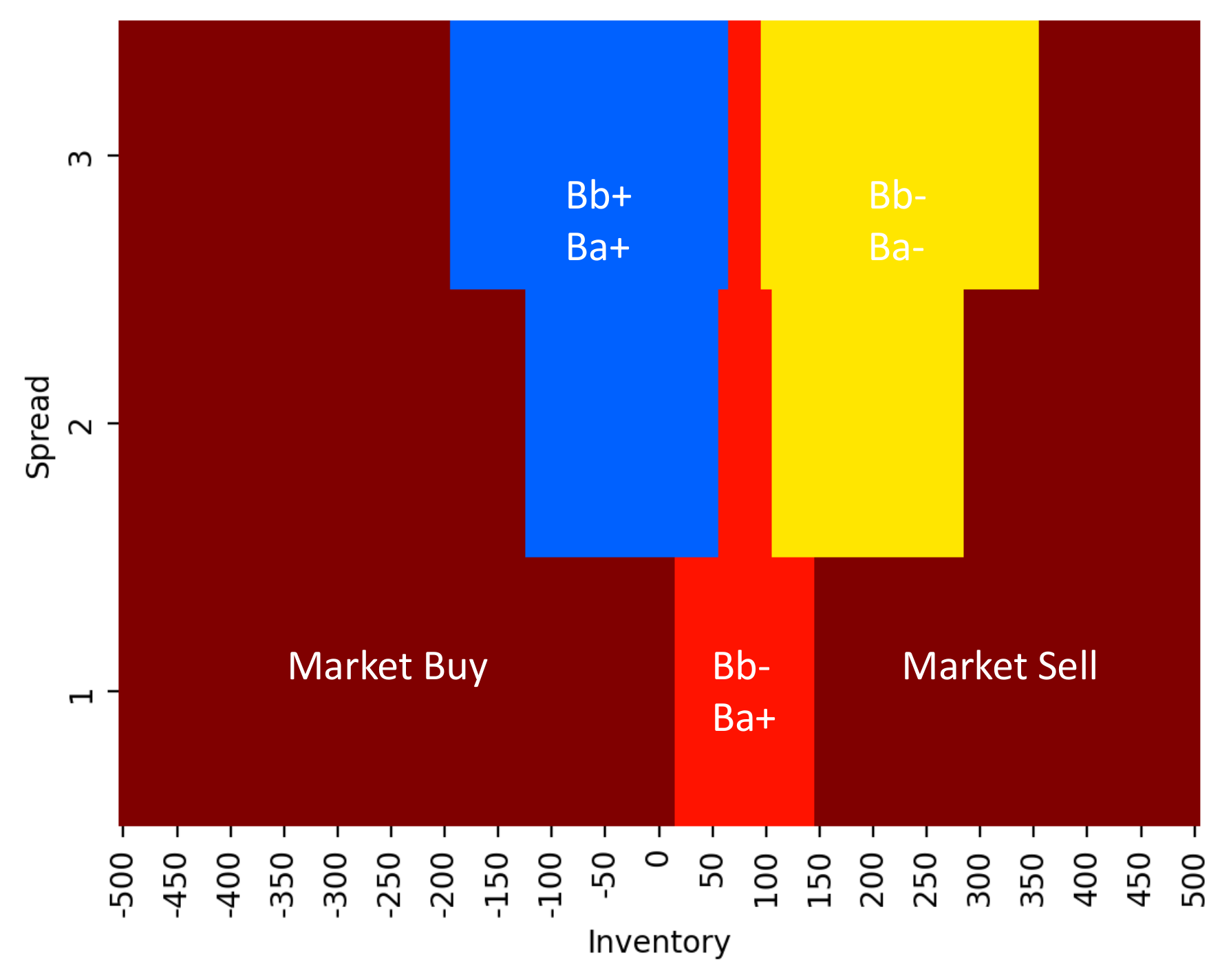}}
 	\caption{Illustrations of the optimal strategies at $t=0$ seconds. Each plot displays the order placement strategy as a function of both inventory and spread. We employ distinct hues to differentiate between the order type and the bid/ask quote of a limit order.}
    \label{fig:order}
\end{figure}

To achieve this, we modify the drift of our baseline strategy while maintaining all other parameters unchanged. We consider three distinct drift levels: $\mu=-0.001$, $0$, and $+0.001$. We then recompute the optimal strategies for each drift level and assess the impact of these drift variations on the strategy. For better understanding, we illustrate the optimal order strategies at time $t=0$ in Figure~\ref{fig:order}. Following this, we backtest these strategies and present the performance results in Table~\ref{table:drift}.

\begin{table}[H]
\centering
\begin{tabular}{|r|c|c|c|}
\hline $\mu$ & Mean of profit & Std of profit & Information Ratio\\
\hline -0.001 & 47.584 & 9.659 & 4.926 \\
\hline 0 & 23.947 & 6.094 & 3.929 \\
\hline 0.001 &47.619 & 9.726 & 4.896 \\
\hline
\end{tabular}
\caption{Profit sensitivity to drift levels}
\label{table:drift}
\end{table}

Figure~\ref{fig:order} demonstrates that when the drift is neutral ($\mu=0$), the order strategy at time t = 0 is well-balanced and symmetric. 
With small initial inventory and low bid-ask spread, if the maker set no restriction on himself or herself to post limit orders on the best bid and ask levels, he or she may choose to post outside the best levels in order to gain larger profits from the spread.   With the increase of the inventory (say net long), if the spread is large thus good opportunity to profit from the spread, the maker choose to take this spread profit opportunity while lower both bid and ask price by one tick in order to enhance the sell opportunity and reduce the buy opportunity.  If the inventory (say net long) is large enough and/or the spread profit can not compensate the inventory risk, the maker will use the market order to get rid of the inventory immediately.  

In the case of positive drift ($\mu=0.001$), where the market maker anticipates an upward movement in price, the maker strategy tends to move up bid and ask quotes and favor a long position to capitalize on the expected increase in price. Conversely, with negative drift ($\mu=-0.001$), the strategy leans towards taking a short position, as the market maker predicts a decrease in stock price.  

Table~\ref{table:drift} presents the performance on three drift levels. In the neutral drift scenario ($\mu=0$), the mean profit and information ratio are both lower compared to the non-neutral drift scenarios ($\mu=0.001, -0.001$). This suggests that when the drift estimation is accurate and aligned with the actual drift in the market, the strategy can generate higher profits and achieve better risk-adjusted performance. As known by the practitioners and pointed out by various researchers \cite{buylowsellhigh2014}, the informed traders has clear target price and usually trade in a directional manner (i.e., price drift), this impose adverse selection risk on the market makers.  If the market is unaware the price drift, significant loss could be resulted in.   On the contrary, an accurate forecast of the price drift by the maker and order posting in advance along the predicted price will effectively enhance the maker's ability to manage the adverse selection risk and gain more profits.

\section{Discussions and Conclusions} \label{sec:disscusion}
This study investigated market making in the Chinese stock market. By simulating the market maker's behavior under various market conditions, our numerical experiments shed light on several important aspects of market making in this market.

First, our results demonstrate the impact of volatility. Specifically, volatility has direct effects on the market maker's inventory. Higher volatility leads to increased inventory risk, making it necessary for the market maker to maintain lower inventory levels in order to mitigate the risk of holding large positions.

Second, our findings indicate that the stamp duty rate is a critical factor in market making. The stamp duty rate has a negative impact on both the profit of the market maker and the liquidity of the market. Moreover, our analysis reveals that the total tax revenue amount is almost a concave function of the stamp tax rate, suggesting that policymakers can use this relationship to set stamp duty rates in a way that maximizes tax revenue without significantly harming market liquidity or the profitability of market makers.

Third, our study emphasizes the significance of considering the impact of price drift on market making strategies. Accurately estimating the drift is crucial for the market maker to optimize their order strategies and manage inventory effectively. In other words, accurate drift estimation leads to increased profits for the market maker.

While the results of this study have important implications for market makers and policymakers in the Chinese stock market, there are some limitations to our research. For example, we assume the mid-quote $P_t$ follows a continuous-state diffusion process, which may not be an ideal fit for the short-term evolution of the mid-quote due to its discreteness resulting from the minimum tick size. Additionally, we relied on certain assumptions in our model and data, such as the independence between the mid-quote $P_t$ and spread $S_t$. Despite these limitations, our findings contribute to the current understanding of market making in the Chinese stock market and offer directions for further research in designing effective market making strategies.

\section{Acknowledgements}
The authors wish to thank Pengcheng Laboratory for the support of the quantitative finance research project.  Thanks to Jimin Han for his help with  data cleaning and the reconstruction of the limit order book.  

\bibliography{ref}

\begin{thebibliography}{10}

\bibitem{avellaneda2008high}
Marco Avellaneda and Sasha Stoikov.
\newblock High-frequency trading in a limit order book.
\newblock {\em Quantitative Finance}, 8(3):217--224, 2008.

\bibitem{Banshortselling2022}
Wolfgang Bessler and Marco Vendrasco.
\newblock The 2020 european short-selling ban and the effects on market
  quality.
\newblock {\em Finance Research Letters}, 42:101886, 2021.

\bibitem{Noise1986}
Fischer Black.
\newblock Noise.
\newblock {\em The Journal of Finance}, 41(3):528--543, 1986.

\bibitem{buylowsellhigh2014}
\'{A}lvaro Cartea, Sebastian Jaimungal, and Jason Ricci.
\newblock Buy low, sell high: A high frequency trading perspective.
\newblock {\em SIAM Journal on Financial Mathematics}, 5(1):415--444, 2014.

\bibitem{CONG201623}
F.~Cong and C.W. Oosterlee.
\newblock Multi-period mean–variance portfolio optimization based on
  monte-carlo simulation.
\newblock {\em Journal of Economic Dynamics and Control}, 64:23--38, 2016.

\bibitem{FORSYTH2011241}
Peter~A. Forsyth.
\newblock A hamilton–jacobi–bellman approach to optimal trade execution.
\newblock {\em Applied Numerical Mathematics}, 61(2):241--265, 2011.

\bibitem{FORSYTH2007}
Peter~A Forsyth and George Labahn.
\newblock Numerical methods for controlled hamilton-jacobi-bellman pdes in
  finance.
\newblock {\em Journal of Computational Finance}, 11(2):1, 2007.

\bibitem{guilbaud2013optimal}
Fabien Guilbaud and Huyen Pham.
\newblock Optimal high-frequency trading with limit and market orders.
\newblock {\em Quantitative Finance}, 13(1):79--94, 2013.

\bibitem{Optmarketmaking2017}
Olivier Guéant.
\newblock Optimal market making.
\newblock {\em Applied Mathematical Finance}, 24(2):112--154, 2017.

\bibitem{inventory2013}
Olivier Guéant, Charles-Albert Lehalle, and Joaquin Fernandez-Tapia.
\newblock Dealing with the inventory risk: a solution to the market making
  problem.
\newblock {\em Mathematics and Financial Economics}, 7(4):477--507, 2013.

\bibitem{Han8505}
Jiequn Han, Arnulf Jentzen, and Weinan E.
\newblock Solving high-dimensional partial differential equations using deep
  learning.
\newblock {\em Proceedings of the National Academy of Sciences},
  115(34):8505--8510, 2018.

\bibitem{HANIF2014429}
Ayub Hanif.
\newblock Chapter 13 - high frequency trading and black box models.
\newblock In Robert Kissell, editor, {\em The Science of Algorithmic Trading
  and Portfolio Management}, pages 429--451. Academic Press, San Diego, 2014.

\bibitem{john2021calculating}
Majnu John and Yihren Wu.
\newblock Calculating infinitesimal generators.
\newblock {\em Journal of Stochastic Analysis}, 2(4):4, 2021.

\bibitem{LI2010448}
Wei Li and Steven~Shuye Wang.
\newblock Daily institutional trades and stock price volatility in a retail
  investor dominated emerging market.
\newblock {\em Journal of Financial Markets}, 13(4):448--474, 2010.

\bibitem{HFTChinaLIU2021}
Wei Liu.
\newblock Can hft profit in chinese stock market?
\newblock {\em Economics Letters}, 209:110115, 2021.

\bibitem{HestonVol2020}
Qing-Qing Yang, Wai-Ki Ching, Jiawen Gu, and Tak-Kuen Siu.
\newblock {Trading strategy with stochastic volatility in a limit order book
  market}.
\newblock {\em Decisions in Economics and Finance}, 43(1):277--301, June 2020.

\bibitem{ChinaMarket2022}
Ruixun Zhang, Chaoyi Zhao, Yufan Chen, Lintong Wu, Yuehao Dai, Ermo Chen,
  Zhiwei Yao, Yihao Zhou, and Lan Wu.
\newblock High-frequency liquidity in the chinese stock market: Measurements,
  patterns, and determinants.
\newblock {\em Available at SSRN}, 2022.

\end{thebibliography}
\bibliographystyle{plain}

\appendix

\begin{appendices}

\section{Mathematical Modelling of a Market Maker}
\label{app:a}
\subsection{The Limit Order Book (LOB)}
As a simplification, the prices of a stock in the LOB  are simply modeled  as the mid-quote and the bid-ask spread.

The mid-quote $P_t$ of the stock price is assumed to follow an exogenous drifted diffusion process:
\begin{align}
dP_t = \mu dt + \sigma d W_t,
\label{eq:p}
\end{align}
where $W_t$ is a standard Brownian motion, $\mu$ is the drift and $\sigma$ is the volatility of the stock price, respectively.

Following \cite{guilbaud2013optimal}, we use an exogenous finite-state continuous process $S_t$ to denote the bid-ask spread of the risk asset at time $t$. With a minimum tick size $\delta$, the spread $S_t$ takes values in $\mathbb{S}=\delta\mathbb{I}_m$ and jumps at random times, where $\mathbb{I}_m = \{1,\cdots,m\}$ and $m\in \mathbb{N}^{+}$ is a constant. To model the jump transitions of $S_t$, two independent processes $N_t$ and $\hat{S}_n$ are introduced. $N_t$ is a Poisson process with a deterministic intensity $\lambda(t)$ to represent the cumulative count of random bid-ask spread jumps by time $t$. $\hat{S}_n$ is a discrete-time Markov chain valued in $\mathbb{S}$ with a probability transition matrix $\mathrm{P}[\hat{S}_{n+1} = j\delta | \hat{S}_n= i\delta] = \rho_{ij}, 1\leq i,j \leq m$, to represent the transition of spread values. Hence, the spread process $S_t$ is characterized by $S_t = \hat{S}_{N_t}, t \geq 0$, which is a continuous-time Markov chain with transition matrix ${R}_S(t) =(r_{ij}(t))_{1\leq i,j \leq m}$, where $r_{ij}(t) = \lambda(t) \rho_{ij}$ for $i\neq j$, and $r_{ii}(t) = - \sum_{j\neq i} r_{ij}(t)$. 
The best-bid and best-ask price are defined as $P_t^b = P_t - \frac{S_t}{2}$ and $P_t^a = P_t + \frac{S_t}{2}$, and $S_t$ and $P_t$ are assumed independent. 

\subsection{The Limit Order Strategies of the Market Maker}

The limit order strategies of the market maker is modeled as a continuous-time predictable control process as follows:
\begin{align}
\alpha_t^{make}=(Q^b_t,Q^a_t,L^b_t,L^a_t), \,\, t\geq 0, \label{eq: make2}
\end{align}
where $L^b_t \in [0,\bar{l}]$ and $L_t^a  \in [0,\bar{l}], \bar{l}>0$ represent the size of the buy and sell limit orders, and $Q^b_t \in \mathbb{Q}^b = \{Bb_-, Bb, Bb_+\}$ and $Q^a_t \in \mathbb{Q}^a = \{Ba_+, Ba, Ba_-\}$ represent the corresponding bid quote and ask quote, respectively. Here we consider three quote levels for buy and sell orders as shown in Figure~\ref{fig:LOB}. For the bid quote $Q_t^b$, $Bb$ denotes the best bid quote $P^b_t$, $Bb_+$ denotes the best bid quote plus one tick at $P^b_t + \delta$, and $Bb_-$ denotes the best bid quote minus one tick at $P^b_t - \delta$. For the ask quote $Q_t^a$, similarly, $Ba$ denotes the best ask quote $P^a_t$, $Ba_-$ denotes the best ask quote minus one tick at $P^a_t - \delta$, and $Bb_+$ denotes the best ask quote plus one tick at $P^a_t + \delta$.

We use $\pi^b(Q_t^b, P_t, S_t)$ and $\pi^a(Q_t^a, P_t, S_t)$ to represent the limit bid and ask order prices at time $t$ taking commission and stamp tax into consideration, which are written as
\begin{align}
\pi^b(q^b,p,s) &=  \begin{cases}
  (p - \frac{s}{2} - \delta)(1+\varepsilon)  & \text{if } q^b =Bb_{-} \\
  (p - \frac{s}{2} )(1+\varepsilon) & \text{if } q^b =Bb \\
  (p - \frac{s}{2} + \delta)(1+\varepsilon) & \text{if }  q^b =Bb_{+}
\end{cases},   \label{eq:pi_b}\\
\pi^a(q^a,p,s) &= \begin{cases}
  (p + \frac{s}{2} - \delta)(1-\varepsilon-\rho)  & \text{if } q^a =Ba_{-} \\
  (p + \frac{s}{2})(1-\varepsilon-\rho) & \text{if } q^a =Ba \\
  (p + \frac{s}{2} + \delta)(1-\varepsilon-\rho) & \text{if } q^a =Ba_{+}
\end{cases}, \label{eq:pi_a}
\end{align}

where $\varepsilon$ and $\rho$ represent the commission rate and stamp tax rate, respectively.

We assume that the limit buy and sell orders of the market maker in question are small orders and fulfilled by the incoming market orders from the counter-parties.  And we further assume that the cumulative execution count of limit buy and sell orders by time $t$ as independent Cox processes $N^b_t$ and $N^a_t$, whose intensities are only related to the quote and the spread as $\lambda^b(Q^b_t,S_t)$ and $\lambda^a(Q^a_t,S_t)$, respectively. According to market characteristics, the following inequalities should hold for all $s \in \mathbb{S}$: $\lambda^b(Bb_{-},s) < \lambda^b(Bb,s) < \lambda^b(Bb_{+},s)$ and $\lambda^a(Ba_+,s) < \lambda^a(Ba,s) < \lambda^a(Ba_{-},s)$. 

Thus, for a limit order strategy $\alpha_t^{make}$, the cash amount $X$ and the number of shares $Y$ held by the market maker follow the equations:
\begin{align}
dY_t &= L^b_t dN^b_t - L^a_t dN^a_t \\
dX_t &= -\pi^b(Q^b_t, P_{t^{-}}, S_{t^{-}}) L^b_t dN^b_t + \pi^a(Q^a_t, P_{t^{-}}, S_{t^{-}}) L^a_t dN^a_t.
\label{eq:xylimit}
\end{align}

\subsection{The Market Order Strategies of the Market Maker}
 Market orders are used when the trader wishes to execute the trade immediately without waiting.  This avoids the non-execution risk of the order at the cost that the transaction may not be fulfilled at a designated or more favored price level. In our study, as a simplification the market orders are assumed sufficiently small in volume and immediately executed at the best price in LOB without a price impact. Thus, the market order strategy is modeled simply as an impulse control:
\begin{align}
  \alpha^{take}=(\tau_n,\zeta_n)_{n\geq 0},
  \label{eq:takeAppex}
\end{align}
where $\tau_n$ is a stopping process that represents the time at which the market maker's $n^{th}$ market order is placed, and $\zeta_n$ is a random variable that represents the size of the $n^{th}$ market order, taking values in the $[-\bar{e},\bar{e}]$, where $\bar{e} > 0$. If $\zeta_n \geq 0$, then the market order buys $\zeta_n$ shares at the current best ask price. If $\zeta_n < 0$, then the market order sells $-\zeta_n$ shares at the current best bid price.

The changes in cash and inventory are thus jump processes, and the changes at time $\tau_n$ can be described by the following equations:
\begin{align}
  Y_{\tau_n} &= Y_{\tau^{-}_n} + \zeta_n,   \\
  X_{\tau_n} &= X_{\tau^{-}_n} - c(\zeta_n, P_{\tau_n}, S_{\tau_n}),
  \label{eq:xymarket}
\end{align}
where
\begin{align}
c(e, p, s) = (e+ \varepsilon |e| + \rho |e| \cdot 1_{\{e<0\}})p + (|e|+  \varepsilon e + \rho e \cdot 1_{\{e<0\}})\frac{s}{2} \label{eq:c}
\end{align}
represents the amount of cash corresponding to an order volume of $e$, a stock mid-quote of $p$, and a spread of $s$, and $\varepsilon$ and $\rho$ represent the commission rate and stamp tax rate, respectively.

\subsection{Optimal Order Strategies}
Following the framework of \cite{guilbaud2013optimal}, over a finite time horizon $T < \infty$, such as within a trading day, the market maker aims to maximize profit from trading through the bid-ask spread, penalize the net inventory during the course, and liquidate the net position at the end time $T$ (without maintaining overnight positions for the avoidance of the inventory risk and overnight capital consumption). The optimal control can therefore be formulated as
\begin{align}
\max_{\alpha=(\alpha^{take}, \alpha^{make})} \mathbb{E}[U(X_T) -\gamma \int_0^{T} g(Y_t)dt],
\end{align}
where the control $\alpha$ must satisfy $Y_T=0$. $U$ is a monotonically increasing reward function, $g$ is a non-negative, convex function, and $\gamma$ is a non-negative penalty constant which expresses the view of the market maker towards the inventory risk aversion. 

\subsection{Value function}\label{app:valuefunc}
The value function $v(t,x,y,p,s)$ represents the maximum expected utility that an investor can obtain by taking a particular control action in a given state at time $t$. This maximum is taken over all possible control actions $\alpha \in \mathcal{A}$, where $\mathcal{A}$ denotes the set of all the limit and market order strategies $\alpha=(\alpha^{take}, \alpha^{make})$. In the previous section, it was required that the position be liquidated at the terminal time $T$, i.e. $Y_T=0$. In order to remove this requirement, a liquidation function $L(x, y, p, s)$ is introduced, and for a given state characterized by the variables $(x,y,p,s)$, it is defined as:
\begin{align}
\begin{split}
L(x, y, p, s)&=x-c(-y, p, s)\\
 &=x+ (y-\varepsilon |y| - \rho |y| \cdot 1_{\{y<0\}})p-(|y|-\varepsilon y - \rho y \cdot 1_{\{y<0\}}) \frac{s}{2} ,
\end{split}
\label{eq: liquidation}
\end{align}
which represents the total amount of cash that an investor would have if they immediately liquidate their entire position at market price.

With the introduction of the liquidation function, the control problem from the previous section can be rewritten as:
\begin{align}
\max_{\alpha=(\alpha^{take}, \alpha^{make})} \mathbb{E}\left[U\left(L\left(X_{T}, Y_{T}, P_{T}, S_{T}\right)\right)-\gamma \int_{0}^{T} g\left(Y_{t}\right) \mathrm{d} t\right]
\end{align}

The value function for this problem can then be defined as:
\begin{align}
v(t,x,y,p,s)=\max_{\alpha \in \mathcal{A}} \mathbb{E}_{t, x,y,p,s}\left[U\left(L\left(X_{T},Y_{T},P_{T}, S_{T}\right)\right)-\gamma \int_{t}^{T} g\left(Y_{u}\right) \mathrm{d} u\right]
\end{align}
where $\mathbb{E}_{t, x,y,p, s}$ denotes the expected value of the process $(X,Y,P,S)$ with initial values $(X_{t^{-}},Y_{t^{-}},P_{t^{-}},S_{t^{-}})=(x,y,p,s)$. Since $s$ is discrete, the value function can be expressed as  $v_i(t,x,y,p)=v(t,x,y,p,i\delta)$. This control problem is a mixed regular/impulse control problem, and can be solved using the dynamic programming method.

For limit order control, given any $q=(q^b,q^a)$ and $l=(l^b,l^a)$, consider the operator
\begin{align}
\begin{split}
 \mathcal{L}^{q, \ell} v(t, x, y, p, s) &=\mathcal{L}_{P} v(t, x, y, p, s)  +{R}_S(t) v(t, x, y, p, s) \\
 &+A^bv(t,x,y,p,s) +A^av(t,x,y,p,s)
\end{split}
\label{eq:lql2}
\end{align}
where 
\begin{align}
\mathcal{L}_Pv(t,x,y,p,s) &= \mu\frac{\partial v}{\partial p}(t,x,y,p,s) + \frac{\sigma^2}{2} \frac{\partial^2 v}{\partial p^2}(t,x,y,p,s),\\ 
R_S(t) v(t, x, y, p, s) &= \sum_{j=1}^{m} r_{i j}(t)[v(t, x, y, p, j \delta)-v(t, x, y, p, i \delta)], \\
A^bv(t,x,y,p,s) &=\lambda^b(q^b,s)\left[v(t,x-\pi^{\mathrm{b}}\left(q^{\mathrm{b}}, p, s\right) \ell^{\mathrm{b}}, y+\ell^{\mathrm{b}},p,s)-v(t,x,y,p,s)\right], \\
A^av(t,x,y,p,s) &=\lambda^a(q^a,s)\left[v(t,x+\pi^{\mathrm{a}}\left(q^{\mathrm{a}}, p, s\right) \ell^{\mathrm{a}}, y-\ell^{\mathrm{a}},p,s)-v(t,x,y,p,s)\right].
\end{align}

The first term in $\mathcal{L}^{q,l}$ represents the infinitesimal generator of the mid-quote process $P$, the second term represents the generator of the continuous-time Markov chain price process $S$, and the last two terms represent the infinitesimal generators of the jump processes caused by the changes in cash and inventory when the limit order $(Q_t,L_t)=(q,l)$ occurs.
~\\

For market order control, consider the impulse operator
\begin{align}
\mathcal{M} v(t, x, y, p, s)=\max_{e \in[-\bar{e}, \bar{e}]} v\left(t, x-c(e, p, s), y+e, p, s\right) .
\label{eq:m}
\end{align}

When combining limit order and market order controls, the dynamic programming equation for this control problem is
\begin{align}
\min \left[-\frac{\partial v_{i}}{\partial t}-\max_{(q, \ell) \in \mathbb{Q}^b \times \mathbb{Q}^a \times[0, \bar{\ell}]^{2}} \mathcal{L}^{q, \ell} v+\gamma g, v-\mathcal{M} v\right]=0,
\end{align}
where the terminal condition is
\begin{align}
  v(T, x, y, p, s)=U(L(x, y, p, s)).
\end{align}

Furthermore, the HJB equation for $v_{i} (1\leq i \leq m)$ reads,
\begin{align} \label{eq:hjbpde}
\begin{split}
 \min \Bigg[ &-\frac{\partial v_{i}}{\partial t}-\mu \frac{\partial  v_{i}}{\partial p} + \frac{\sigma^2}{2} \frac{\partial^2  v_{i}}{\partial p^2}-\sum_{j=1}^{m} r_{i j}(t)\left[v_{j}(t, x, y, p)-v_{i}(t, x, y, p)\right] \\
 &-\max_{\left(q^{\mathrm{b}}, \ell^{\mathrm{b}}\right) \in \mathcal{Q}_{i}^{\mathrm{b}} \times[0, \bar{\ell}]} \lambda^{\mathrm{b}}\left(q^{\mathrm{b}},i\delta\right)\left[v_{i}\left(t, x-\pi^{\mathrm{b}}\left(q^{\mathrm{b}}, p, i\delta\right) \ell^{\mathrm{b}}, y+\ell^{\mathrm{b}}, p\right) -v_{i}(t, x, y, p)\right] \\
 &-\max_{\left(q^{\mathrm{a}}, \ell^{\mathrm{a}}\right) \in \mathcal{Q}_{i}^{\mathrm{a}} \times[0, \bar{\ell}]} \lambda^{\mathrm{a}}\left(q^{\mathrm{a}},i\delta \right)\left[v_{i}\left(t, x+\pi^{\mathrm{a}}\left(q^{\mathrm{a}}, p, i\delta\right) \ell^{\mathrm{a}}, y-\ell^{\mathrm{a}}, p\right)-v_{i}(t, x, y, p)\right] \\
 &+\gamma g(y), \left.v_{i}(t, x, y, p)-\max_{e \in[-\bar{e}, \bar{e}]} v_{i}\left(t, x-c(e, p, i\delta), y+e, p\right)\right]=0,   
\end{split}
\end{align}
where function $\pi^a,\pi^b$ and $c$ are defined in Equation~\eqref{eq:pi_a},\eqref{eq:pi_b},\eqref{eq:c}.

\end{appendices}
 
\end{document}